\documentclass[letterpaper]{aa}

\usepackage[final]{epsfig}

\usepackage{times}
\usepackage{latexsym}
\usepackage{amssymb}

\newcommand{\kmps}{\ensuremath{\mathrm{km\,s^{-1}}}}
\newcommand{\percc}{\ensuremath{\mathrm{cm^{-3}}}}
\newcommand{\Msun}{\ensuremath{\mathrm{M}_\odot}}
\newcommand{\kkms}{\ensuremath{\mathrm{K\,km\,s^{-1}}}}
\newcommand{\hour}{\ensuremath{^\mathrm{h}}}
\newcommand{\minute}{\ensuremath{^\mathrm{m}}}
\newcommand{\second}{\ensuremath{^\mathrm{s}}}
\newcommand{\twco}{\ensuremath{\mathrm{^{12}CO}}}
\newcommand{\thco}{\ensuremath{\mathrm{^{13}CO}}}
\newcommand{\ceio}{\ensuremath{\mathrm{C^{18}O}}}
\newcommand{\ceiot}{{\hbox {\ensuremath{\mathrm{C^{18}O}~(J=1-0)} }\ }}
\newcommand{\thcot}{{\hbox {\ensuremath{\mathrm{^{13}CO}~(J=1-0)} }\ }}
\newcommand{\thcotp}{{\hbox {\ensuremath{\mathrm{^{13}CO}~(J=1-0)} }}}
\newcommand{\csot}{{\hbox  {\ensuremath{\mathrm{C^{17}O}~(J=1-0)} }\ }}
\newcommand{\twcot}{{\hbox {\ensuremath{\mathrm{^{12}CO}~(J=1-0)} }\ }}

\begin{document}
   \title{Clumpy filaments of the Chamaeleon~{\sc i} cloud:\\
          C$^{18}$O mapping with the SEST\thanks{Based 
on observations collected at the European Southern Observatory,  La Silla, Chile
}}

   \author{L. K. Haikala
          \inst{1,2}
          \and
          J. Harju\inst{2}
          \and K. Mattila\inst{2}
          \and M. Toriseva\inst{2}
          }

   \offprints{L. Haikala }

   \institute{Swedish-ESO Submillimetre Telescope, 
European Southern Observatory, Casilla 19001, Santiago, Chile\\
\email{haikala@astro.helsinki.fi}              
         \and
Observatory, PO Box 14, FI-00014 University of Helsinki, Finland
             }

   \date{Received ; accepted }

\abstract{The Chamaeleon~{\sc i} dark cloud (Cha~{\sc i}) has been
mapped in \ceiot\ with an angular resolution of $1\arcmin$ using the
SEST telescope.  The large scale structures previously observed with lower
spatial resolution in the
cloud turn into a network of clumpy filaments.  The automatic
Clumpfind routine developed by \cite{williams1994} is used to identify
individual clumps in a consistent way. Altogether 71 clumps were found and 
the total mass of these clumps is 230 $\Msun$. The dense 'cores' detected
with the NANTEN telescope (\cite{mizuno1999}) and the very cold cores
detected in the ISOPHOT serendipity survey (\cite{toth2000}) form
parts of these filaments but decompose into numerous 'clumps'.  The
filaments are preferentially oriented at right angles to the
large-scale magnetic field in the region.  We discuss the cloud
structure, the physical characteristics of the clumps and the
distribution of young stars.  The observed clump mass spectrum is
compared with the predictions of the turbulent fragmentation model of
\cite{padoan2002}. An agreement is found if fragmentation has been
driven by very large-scale hypersonic turbulence, and if it has had
time to dissipate into modestly supersonic turbulence in the
interclump gas by the present time. According to numerical
simulations, large-scale turbulence should have resulted in
filamentary structures as seen in Cha~{\sc i}. The well-oriented
magnetic field does not, however, support this picture, but
suggest magnetically steered large-scale collapse. The origin of
filaments and clumps in Cha~{\sc i} is thus controversial.  A possible
solution is that the characterization of the driving turbulence fails
and in fact different processes have been effective on small and large
scales in this cloud.

   \keywords{ ISM:clouds -- ISM:molecules --  ISM:structure -- 
              individual objects: Chamaeleon I}
}
\titlerunning  {\ceiot\ in Chamaeleon I}

   \maketitle
%

\section{Introduction}

The dark cloud Cha~{\sc i} is one of the nearest
low-mass star forming regions. 
The distance 150 pc to the cloud has been recently
determined using the Hipparcos satellite data by \cite{knude1998}.
The elongated cloud with the apparent dimensions
$\sim 0.5\degr$ by $\sim 2 \degr$ is easily recognized on the ESO/SRC
sky survey plates. A visual extinction map based on star counts on the
ESO/SRC sky survey plates is presented in \cite{toriseva1985} and a
near-IR extinction map from DENIS $IJK$ star counts by
\cite{cambresy1997}.  Several signposts of past and present star
formation are observed in the cloud.  Three visually bright reflection
nebulae (Ced 110, Ced 111 and Ced 112; \cite{cederblad1946}) and an
infrared reflection nebula (the IRN) are associated with Cha~{\sc i}.

More than hundred pre-main-sequence stars have been found in the
cloud.  Besides clusters around the visible reflection nebulae, these
objects are scattered around its western mid part and the northern
part.  Evidence for ongoing star formation near Ced 110 and Ced 112
is provided by observed molecular outflows (\cite{mattila1989},
\cite{prusti1991}) and two mm continuum sources
(\cite{reipurth1996}). The research on Cha~{\sc i} until 1990 is
reviewed by \cite{schwartz1991}. A short summary of later large scale
studies is given in the following.

A search for new young stellar objects in Cha~{\sc i} has been
conducted by \cite{cambresy1998} using DENIS $IJK$ photometric
data. Most sources were distributed in a similar way to the known
member candidates in the central and northern parts of the cloud. Only
few young star candidates were found in the southern part. A deeper
$JHK$ study by \cite{gomez2001} confirmed these results.  New
candidate member stars have been found using $L$-band photometry
(\cite{kenyon2001}) and $JHK$ variability studies
(\cite{carpenter2002}).  Using ISOCAM observations \cite{persi2000}
found clusters of sources with mid-IR excess around the three
reflection nebulae.  A number of sources with mid-IR excess is seen
also in the very eastern and western parts of the cloud but only one
in the dense region southwest of Ced 111.

The multicolour IRAS images (\cite{boulanger1990};
\cite{boulanger1998}) outline the dense parts of Cha~{\sc i}.  FIR
ISOPHOT observations of the southern part are reported in Haikala et
al 1998.  Besides the emission from the strong IR point sources in the
reflection nebulae Ced 110 and 111 and the IRN, extended FIR emission
is seen to coincide with the extinction features mapped by
\cite{toriseva1985} and \cite{cambresy1997}.  In particular two
extended dust emission maxima are seen southwest of Ced 111. These
dust emission maxima were detected also by the ISOPHOT 170 $\mu$m
Serendipity Survey observations of Cha~{\sc i}
(\cite{toth2000}). However, up to now no FIR point sources have been
detected in these maxima.

Compared to the major star formation regions in the northern sky,
radio molecular line observations of the Chamaeleon region have been
few.  The first large-scale radio molecular line mappings of the cloud
were made with the Parkes 64-m telescope in the 6 and 18 cm
transitions of H{$_2$CO and OH, respectively (\cite{toriseva1985a}).
The 1.2-m Columbia Southern Millimeter-Wave Telescope at Cerro Tololo
was used to map Cha~{\sc i} in the \ceiot (\cite{toriseva1990}) and
the \twcot (\cite{boulanger1998}).  Due to the small size of the
telescope, the spatial resolution was moderate ($9\arcmin$ HPBW).
First observations in the $1\arcmin$ resolution range were obtained
when the SEST telescope started observations in 1988. As the beam of
the telescope is $45\arcsec$ to $55\arcsec$ at 3 mm, only small areas
were mapped (\cite{mattila1989}). The Nagoya 4-m survey telescope,
NANTEN, located at the Las Campanas Observatory, was used for mapping
the Chamaeleon region in the \twco, \thco\ and \ceiot transitions
(\cite{mizuno2001}; \cite{mizuno1999}; \cite{mizuno1998}).  The beam
size of the NANTEN telescope is 2.7 arcminutes but the observing grid
was $8\arcmin$, $4\arcmin$ and $2\arcmin$, respectively.
 
This paper reports the mapping of the dense parts of the
Cha~{\sc i} cloud in the \ceiot line with an angular resolution
of $1\arcmin$.  These observations enable more detailed studies of the
cloud structure than has been possible by using optical or NIR star
counts (statistical limitations) or by the smaller size radio
telescopes (diffraction limited beam sizes of several arc
minutes). The relatively high frequency resolution of 43 kHz
(corresponding to $\sim0.12$ $\kmps$ at 109 GHz) makes it
possible to extend the analysis also into the velocity space.  An
automated approach to data analysis, the "Clumpfind" routine developed
by \cite{williams1994}, has been used to analyze the $(x,y,v)$ data
cube derived from the observations.

The observations and the data reduction are described in Sect.~2.
In Sect.~3 we describe the cloud structure and results of the
clump finding procedure, and discuss the new results in the light of
previous observations concerning the signposts of star
formation and magnetic fields in the cloud.  A comparison with the
NANTEN cores and clumps identified in this paper is presented in Sect.~4, and
notes of some individual regions of special interest are presented in
Sect.~5. In Sect.~6 we discuss the clump stability and mass
spectrum. Finally, in Sect.~7 we summarize our conclusions.}


\section{Observations}
The observations were conducted during various observing runs from
1988 to June 1996 using the Swedish-ESO-Submillimetre-Telescope SEST
on La Silla observatory in Chile.  A 3~mm dual polarization single
sideband (SSB) Schottky receiver in the frequency switching mode (with
a throw of 7\,MHz) was used to observe the $J=1-0$ transitions of
C$^{18}$O, $^{13}$CO and C$^{17}$O at 109, 110 and 111\,GHz,
respectively.  The SEST high resolution, 2000 channel acousto-optical
spectrometer with an 86\,MHz bandwidth and a channel width of 43\,kHz
was split into two bands to measure the both polarizations
simultaneously. The last C$^{18}$O observations (1995,1996) were done
using the SEST 3mm SSB SIS receiver.

Calibration was achieved by the chopper wheel method.  As the
observations are spread over many observing runs with different
weather conditions the system temperatures varied accordingly.
Typical values for the effective SSB system temperatures outside the
atmosphere ranged from 350~K to 450~K (for the Schotky receiver) and
from 200~K to 300~K (SIS receiver) for the C$^{18}$O measurements, and
from 300~K to 350~K for the $^{13}$CO measurements.

During the different runs the same reference position was always
observed to monitor the calibration.  Pointing was checked every 2-3
hours towards the nearby SiO maser source U~Men. We estimate the
pointing accuracy to have been better than $10^{\prime\prime}$ during
the observing runs. The focusing was done using a strong SiO maser.

{\bf C$^{18}$O}~~ Altogether 1836 positions in the cloud were observed
using a map step size of $1\arcmin$ and integration time of 60
seconds. The observations cover $\sim$0.6 square degrees.  The map is
not fully sampled since the SEST beamsize at this frequency is
$45\arcsec$.  We believe however that the clump detection is mainly
limited by the noise level reached, and not by the slight
undersampling. The $1\arcmin$ step size corresponds to $\sim0.044$ pc
at the distance of 150 pc to the cloud. The $(0,0)$ position was
arbitrarily chosen to coincide with star T39 near the cloud centre
(J2000.0: $11\hour 09\minute13\fs3$, $-77\degr 29\arcmin 12\arcsec$).
The C$^{18}$O observations of the Ced 110 and 112 regions
have already been reported in \cite{mattila1989}.

The median RMS noise of the C$^{18}$O spectra, after folding and
baseline fit (typically second order baseline was used), is 0.1 K.
Because of different observing conditions the noise goes occasionally
up, and the maximum level is 0.22 K. For 90\% of the spectra the RMS
noise is smaller than 0.15 K.

\begin{figure*} \centering \includegraphics[bb=10 0 763 800,
   width=20.8cm]{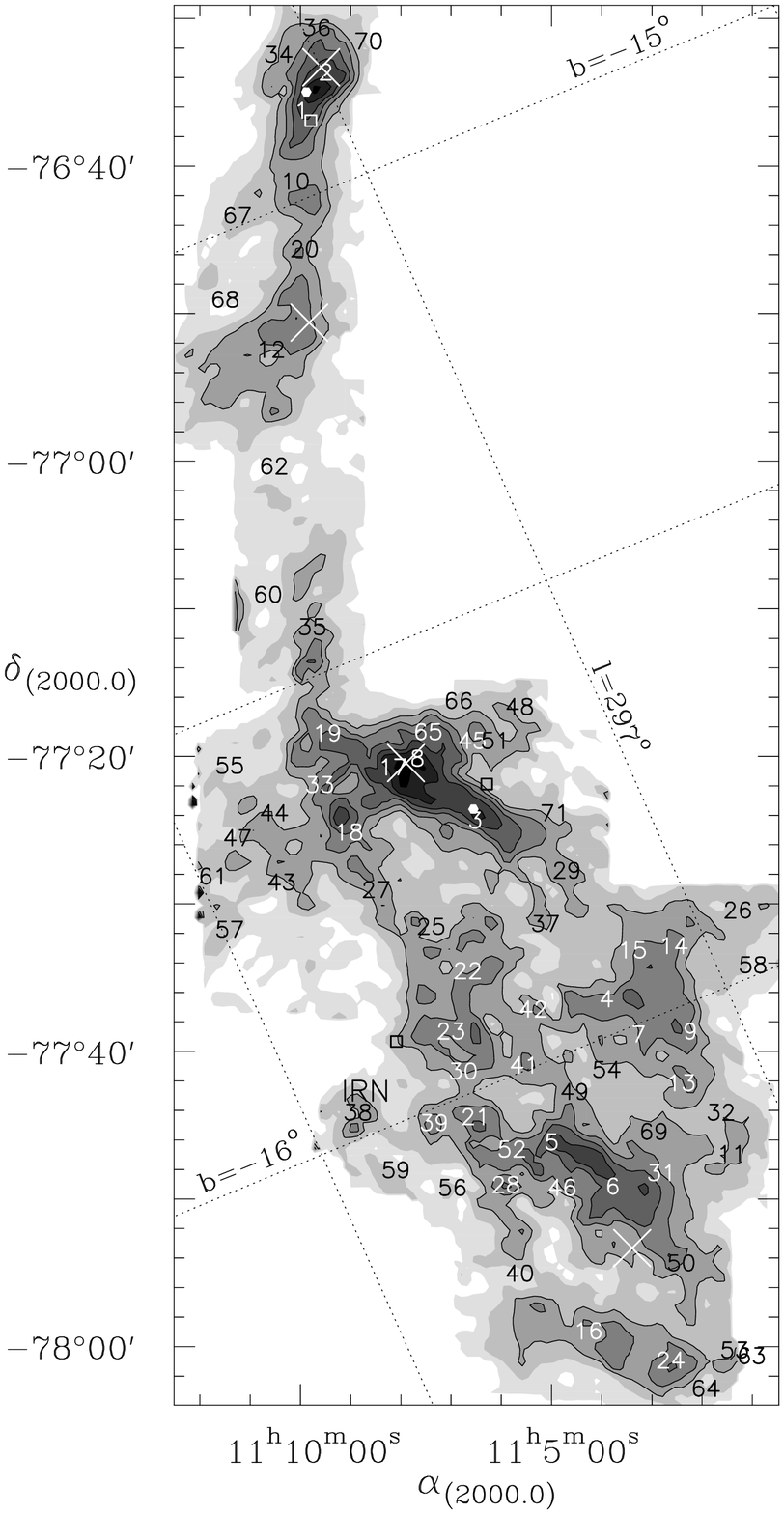}
   \caption{Integrated C$^{18}$O   T$_{\rm A}^*$ intensity map over the LSR 
   velocity range    $2.9 - 5.9 \kmps$ of the Chamaeleon~{\sc i}
   cloud.  The lowest  greyscale level is $0.2\, \kkms$  ~and the stepsize 
   is $0.3\,   \kkms$. The lowest contour level is $0.8\, \kkms$. 
   The crosses indicate the  positions of \cite{mizuno1999}
   C$^{18}$O cores 3,4,6 and 5 (from north to south).
   The locations of the reflection nebulae Ced 112, Ced 110
   and Ced 111  (from north to south)
   are indicated with open squares. The locations of the
   Class 0 protostar candidates Cha-MMS1 and Cha-MMS2 are marked with
   filled circles.  'IRN' marks the location of the Infrared
   Reflection Nebula.The numbers identify the clumps found in the cloud
   (see Table.~\ref{table:clumps}).}
   \label{figure:chai_c18o}
   \end{figure*}

\begin{figure*} \centering \includegraphics[
   width=23.5cm,angle=270, clip]{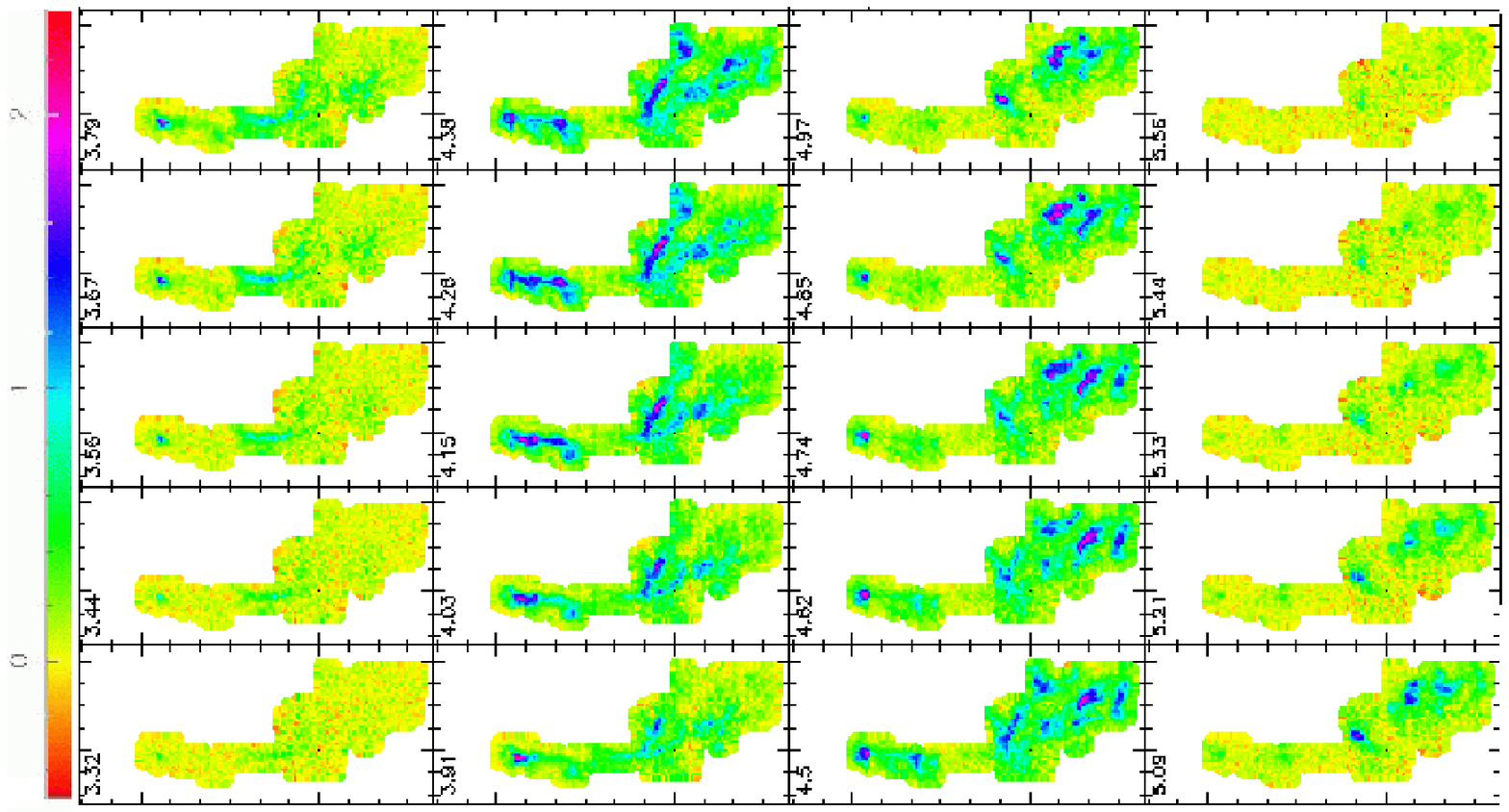}
   \caption{Maps of the C$^{18}$O  T$_{\rm A}^*$ intensity in each of the 
    central channels  in the Chamaeleon~{\sc i} cloud.  The LSR velocity 
   in each panel is indicated in the upper corner of the  panel.
   The highest intensity in the figure is $\sim2.2$ K
   } 
   \label{figure:chai_c18o_channel}
   \end{figure*}

The integrated \ceiot intensity map shown in
Fig.~\ref{figure:chai_c18o} serves as a finding chart for the
identified clumps (see Section 3) and some prominent objects in the
cloud.  The locations of the cores identified by \cite{mizuno1999},
and visible reflection nebulae Ced 110, Ced 111, and Ced 112, and the
IRN are indicated. Also shown are the locations of two Class 0 sources
(Cha-MMS1 and Cha-MMS2) detected by \cite{reipurth1996}.  A velocity
channel map over 20 channels of 0.12 $\kmps$ in width is shown
Fig.~\ref{figure:chai_c18o_channel}.  The velocities are indicated in
the top of each panel.

%

%
%

%

{\bf $^{13}$CO}~~ The central and southern parts of Cha~{\sc i} were
observed in \thcotp.  Observing was done in a similar manner as the
C$^{18}$O observations but with a $2\arcmin$ map step size.  The
number of observed positions is 296 and they cover an area of
$\sim$0.25 square degrees.  The integrated T$_{\rm A}^*$ map is shown
in Fig.~\ref{figure:chai_13co}.  The locations of Ced 110, Ced 111,
and the IRN are also indicated.

{\bf C$^{17}$O}~~ Pointed \csot
observations were made towards a few C$^{18}$O maxima. The integration
time was 20 minutes, and the frequency throw was set to 12 MHz due to
the hyperfine structure of the line.  Longer C$^{18}$O integrations
were made towards the same positions.  In Fig.\ \ref{figure:c17o+c18o}
we present   C$^{17}$O and C$^{18}$O lines observed in three
positions in the cloud.

\section{The cloud structure}


   \begin{figure}  \centering   \includegraphics[bb=25 55 550 550,
   width=10cm,angle=270]{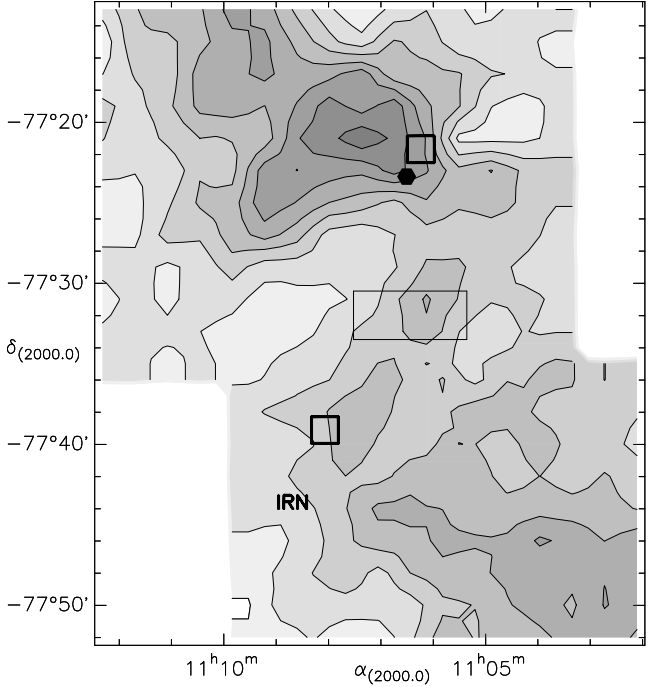}
   \caption{The integrated $^{13}$CO$(J=1-0)$ T$_{\rm A}^*$ intensity
   map of the central part of Cha~{\sc i}. The lowest greyscale level
   is $2.2\, \kkms$ and the stepsize is $0.6\, \kkms$.  The markers
   are the same as in Fig.~\ref{figure:chai_c18o}.  The box in the
   centre of the figure outlines the area displayed in
   Fig.~\ref{figure:spectra}.
   } 
   \label{figure:chai_13co}
   \end{figure}
%

%
   \begin{figure}
   \includegraphics[bb=200 20 550 590,
   width=4cm,angle=270]{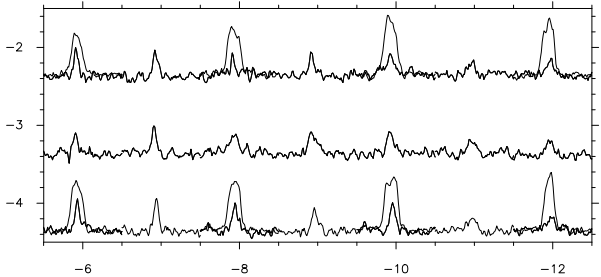}
   \caption{Sample $^{13}$CO and C$^{18}$O $J=1-0$ spectra observed in
   a 6$\arcmin$ by 2$\arcmin$ region indicated in
   Fig.~\ref{figure:chai_13co}.  The maximum $^{13}$CO T$_A^*$ in the
   figure is 3K. The offsets relative to the map centre position
   (R.A. $= 11\hour 09\minute13\fs3$, Dec. $= -77\degr 29\arcmin
   12\arcsec$) in arc minutes are shown on the axes.}
   \label{figure:spectra}
   \end{figure}
%

%

   \begin{figure}   \centering   \includegraphics[bb=160 20 540 360,
    width=9cm, angle=270]{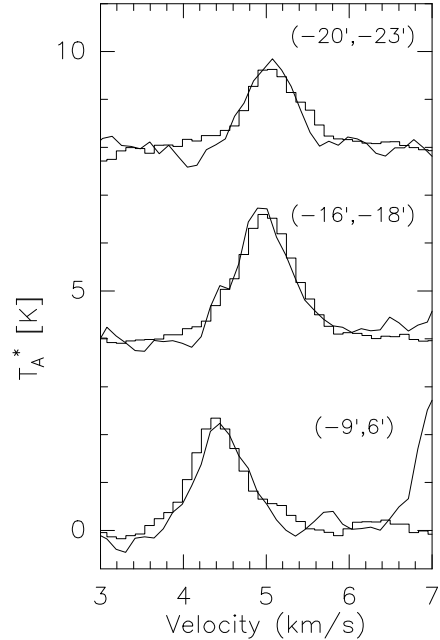}
   \caption{A comparison of the \ceiot\ (histogram) and \csot,
    $F=5/2-5/2$ line (continuous line) intensities in selected \ceio\
    maxima.  The C$^{17}$O intensity has been multiplied by 9, which 
    corresponds to the expected intensity ratio of the \ceiot\ line to this
    C$^{17}$O hyperfine component in the optically thin case.}
   \label{figure:c17o+c18o} 
   \end{figure}
%

The integrated \ceio\ and \thcot intensity maps presented in
Fig.~\ref{figure:chai_c18o} and Fig.~\ref{figure:chai_13co} show
similar features on a large scale.  As the $^{13}$CO mapping was done
using 2$\arcmin$ step size it can not show the cloud structure in
smaller detail. A closer look reveals significant differences between
\ceio \ and \thco . For example, the extended C$^{18}$O maximum near 
~$11\hour03\minute30\second$, $-77\degr 37\arcmin$ is not visible in
the $^{13}$CO map. Another
example is given in Fig.~\ref{figure:spectra}: The intensities of the 
C$^{18}$O and \thcot\ lines  towards the shown positions do
not correlate with each other.  The reason for this failure is that the
$^{13}$CO line becomes optically thick in the direction of a high
column density.  The case of $^{12}$CO is even worse as it is clearly
strongly self absorbed in the line centre almost everywhere (see eg.
Figures 3 and 19 in \cite{mattila1989}).

The comparison of the rare CO isotopomer C$^{17}$O and C$^{18}$O lines
towards three positions in the cloud is shown in Fig.\
\ref{figure:c17o+c18o}.  The two transitions are well correlated and
their intensity ratio (when the three \csot\ components are added)
is 3.6, which corresponds to the abundance ratio in the local ISM
(\cite{wilson1994}). This comparison shows that both lines are likely to be
optically thin and that the C$^{18}$O($J=1-0$) line can be used to
trace the total CO column density distribution.

A conspicuous feature in the C$^{18}$O maps is the presence of several
clumpy filaments. The most prominent of these is the northern arm, a
40$\arcmin$ long north-south oriented structure reaching down from Ced
112.  In the centre and in the south, there are three almost parallel
northeast-southwest oriented filaments, and an arc almost $40\arcmin$
in length, stretching from the neighbourhood of the IRN towards the
northern arm. When investigated in detail these larger entities break
up into smaller clumps. The notable northeast-southwest oriented
filaments break into small, often north-south oriented, clumps.

\subsection{Clump identification}

We have analyzed the small scale structure seen in the C$^{18}$O map
by using the automatic routine ``Clumpfind'' developed by
\cite{williams1994}, which assumes that intensity maxima in the
spectra correspond to localized density enhancements or 'clumps'. The
C$^{18}$O($J=1-0$) spectral line map was interpolated to a
three-dimensional $T_{\rm A}^*(x,y,v)$ data cube, where the pixel size
was $30 \arcsec$ in the position coordinates (one half of the map grid
spacing), and one spectrometer channel width ($0.12 \kmps$) in the
radial velocity.  The C$^{18}$O emission from the cloud is confined to
the velocity range $2.7 - 6.1$ \kmps\ and only the central 29 channels
corresponding to this range were included to the data cube. The maximum
dimensions of the map in R.A. and Dec. directions are $41\arcmin$ and
$96\arcmin$, respectively, and the dimensions of the data cube are
$82\times192\times29$.
\begin{table*}
  \caption[]{Clump properties.}
   \label{table:clumps}
   {\tiny
\begin{tabular}{rccrrccccrc}
No. & $l$ & $b$ &  
 \multicolumn{1}{c}{R.A. (2000)} & \multicolumn{1}{c}{Dec. (2000)} & 
 $V_{\rm LSR}$ &  $\Delta V$ & Radius & $N({\rm C^{18}O})$ & \multicolumn{1}{c}{Mass} & 
 $|E_{\rm p}|/E_{\rm k}$ \\
   & ($\degr$) & ($\degr$)  & 
        ($\hour:\minute:\second$) & $(\degr:\arcmin:\arcsec$) & 
        ($\kmps$) & ($\kmps$) & (pc) & ($10^{15} \,{\rm cm}^{-2}$)& 
     \multicolumn{1}{c}{($\Msun$)} &  \\ \hline
  1  & 297.04 & -14.91 & 11:10:03.8& -76:36:18& 3.8 &  0.6 & 0.15&  1.9&   5.1  ~~(0.9)&  0.9\\ 
  2  & 297.00 & -14.88 & 11:09:34.5& -76:33:43& 4.6 &  0.6 & 0.14&  2.0&   5.2  ~~(0.8)&  1.0\\
  3  & 297.18 & -15.72 & 11:06:30.0& -77:24:18& 4.3 &  0.7 & 0.19&  2.1&  11.7  ~~(2.2)&  1.1\\
  4  & 297.14 & -15.96 & 11:03:50.3& -77:36:24& 4.9 &  0.6 & 0.12&  1.6&   4.5  ~~(0.8)&  0.8\\
  5  & 297.26 & -16.09 & 11:04:54.9& -77:46:08& 4.6 &  0.6 & 0.12&  1.7&   4.0  ~~(0.7)&  0.7\\
  6  & 297.22 & -16.16 & 11:03:40.8& -77:49:04& 4.5 &  0.6 & 0.17&  2.0&  10.1  ~~(1.6)&  1.2\\
  7  & 297.12 & -16.01 & 11:03:09.9& -77:38:44& 4.9 &  0.5 & 0.11&  1.4&   3.7  ~~(0.6)&  0.8\\
  8  & 297.22 & -15.63 & 11:07:39.6& -77:20:09& 5.0 &  0.7 & 0.17&  1.6&   6.9  ~~(1.3)&  0.8\\
  9  & 297.06 & -16.03 & 11:02:07.4& -77:38:34& 4.9 &  0.5 & 0.11&  1.3&   2.8  ~~(0.5)&  0.8\\
 10  & 297.08 & -14.98 & 11:10:10.7& -76:41:05& 4.1 &  0.5 & 0.17&  1.3&   5.9  ~~(1.2)&  1.1\\
 11  & 297.08 & -16.18 & 11:01:17.2& -77:46:52& 4.9 &  0.5 & 0.13&  1.0&   2.6  ~~(0.6)&  0.6\\
 12  & 297.19 & -15.14 & 11:10:40.1& -76:52:31& 4.3 &  0.6 & 0.22&  1.4&  11.6  ~~(2.4)&  1.0\\
 13  & 297.10 & -16.08 & 11:02:18.0& -77:42:03& 4.7 &  0.5 & 0.11&  1.3&   2.7  ~~(0.5)&  0.8\\
 14  & 297.04 & -15.94 & 11:02:28.6& -77:32:44& 4.4 &  0.5 & 0.17&  1.1&   5.8  ~~(1.1)&  1.0\\
 15  & 297.08 & -15.93 & 11:03:17.7& -77:33:05& 4.7 &  0.5 & 0.14&  1.0&   2.6  ~~(0.5)&  0.5\\
 16  & 297.31 & -16.30 & 11:04:09.1& -77:58:56& 4.6 &  0.6 & 0.21&  1.2&   9.1  ~~(2.2)&  0.9\\
 17  & 297.25 & -15.63 & 11:08:09.5& -77:20:46& 4.3 &  0.7 & 0.14&  1.5&   5.7  ~~(1.0)&  0.7\\
 18  & 297.32 & -15.68 & 11:09:02.2& -77:25:10& 4.0 &  0.9 & 0.14&  2.0&   5.4  ~~(1.3)&  0.5\\
 19  & 297.30 & -15.57 & 11:09:29.1& -77:18:32& 4.1 &  0.8 & 0.14&  1.7&   5.5  ~~(1.2)&  0.6\\
 20  & 297.10 & -15.05 & 11:10:00.2& -76:45:44& 4.3 &  0.5 & 0.13&  1.3&   2.9  ~~(0.7)&  0.6\\
 21  & 297.33 & -16.03 & 11:06:26.9& -77:44:26& 4.5 &  0.6 & 0.14&  1.4&   3.4  ~~(0.8)&  0.5\\
 22  & 297.26 & -15.87 & 11:06:37.6& -77:34:30& 4.5 &  0.8 & 0.16&  1.4&   6.7  ~~(1.5)&  0.6\\
 23  & 297.31 & -15.93 & 11:06:58.4& -77:38:41& 4.1 &  0.7 & 0.15&  1.5&   5.2  ~~(1.1)&  0.7\\
 24  & 297.25 & -16.36 & 11:02:29.7& -78:00:55& 4.7 &  0.7 & 0.13&  1.4&   4.2  ~~(1.0)&  0.5\\
 25  & 297.28 & -15.81 & 11:07:22.9& -77:31:31& 4.1 &  0.6 & 0.16&  1.2&   4.5  ~~(1.1)&  0.6\\
 26  & 296.96 & -15.93 & 11:01:11.8& -77:30:23& 4.3 &  0.4 & 0.12&  0.7&   2.7  ~~(0.6)&  0.6\\
 27  & 297.32 & -15.75 & 11:08:28.9& -77:29:08& 4.0 &  0.6 & 0.14&  1.1&   3.4  ~~(0.9)&  0.6\\
 28  & 297.33 & -16.11 & 11:05:50.8& -77:49:02& 4.4 &  0.7 & 0.12&  1.4&   3.1  ~~(0.7)&  0.5\\
 29  & 297.11 & -15.81 & 11:04:38.7& -77:27:44& 4.3 &  0.6 & 0.17&  0.9&   4.7  ~~(1.2)&  0.6\\
 30  & 297.32 & -15.98 & 11:06:42.1& -77:41:22& 3.9 &  0.5 & 0.12&  1.0&   1.9  ~~(0.5)&  0.5\\
 31  & 297.16 & -16.17 & 11:02:41.1& -77:48:10& 5.1 &  0.6 & 0.17&  0.9&   5.4  ~~(1.2)&  0.7\\
 32  & 297.07 & -16.13 & 11:01:30.6& -77:44:03& 4.6 &  0.4 & 0.09&  0.7&   1.1  ~~(0.3)&  0.4\\
 33  & 297.33 & -15.62 & 11:09:37.9& -77:21:59& 4.1 &  0.9 & 0.11&  1.6&   3.2  ~~(0.7)&  0.4\\
 34  & 297.04 & -14.84 & 11:10:32.6& -76:32:33& 4.6 &  0.6 & 0.08&  0.9&   1.2  ~~(0.3)&  0.3\\
 35  & 297.27 & -15.45 & 11:09:47.2& -77:11:19& 3.7 &  0.6 & 0.19&  1.2&   5.4  ~~(1.4)&  0.6\\
 36  & 296.99 & -14.83 & 11:09:45.2& -76:30:40& 4.5 &  0.7 & 0.10&  1.1&   1.8  ~~(0.4)&  0.4\\
 37  & 297.16 & -15.86 & 11:05:04.0& -77:31:19& 5.0 &  0.7 & 0.16&  0.6&   3.2  ~~(1.0)&  0.4\\
 38  & 297.44 & -15.97 & 11:08:49.0& -77:44:13& 4.3 &  0.6 & 0.14&  1.0&   2.6  ~~(0.8)&  0.5\\
 39  & 297.37 & -16.01 & 11:07:17.6& -77:44:50& 3.9 &  0.6 & 0.13&  1.1&   2.3  ~~(0.6)&  0.4\\
 40  & 297.36 & -16.21 & 11:05:33.2& -77:55:02& 4.4 &  0.7 & 0.18&  1.0&   4.2  ~~(1.3)&  0.4\\
 41  & 297.25 & -16.00 & 11:05:28.6& -77:40:56& 4.3 &  0.8 & 0.10&  1.3&   2.2  ~~(0.6)&  0.3\\
 42  & 297.22 & -15.94 & 11:05:18.6& -77:37:08& 4.4 &  0.9 & 0.13&  0.7&   3.3  ~~(0.9)&  0.3\\
 43  & 297.42 & -15.70 & 11:10:23.6& -77:28:35& 4.3 &  0.7 & 0.15&  0.6&   3.3  ~~(1.0)&  0.4\\
 44  & 297.39 & -15.63 & 11:10:33.0& -77:23:57& 4.5 &  0.7 & 0.11&  1.0&   2.3  ~~(0.6)&  0.4\\
 45  & 297.15 & -15.64 & 11:06:32.8& -77:19:00& 4.5 &  0.7 & 0.11&  0.8&   2.1  ~~(0.5)&  0.3\\
 46  & 297.27 & -16.14 & 11:04:41.7& -77:49:14& 4.0 &  0.5 & 0.12&  0.7&   1.8  ~~(0.5)&  0.4\\
 47  & 297.44 & -15.64 & 11:11:17.4& -77:25:40& 4.5 &  0.7 & 0.09&  1.0&   1.3  ~~(0.4)&  0.3\\
 48  & 297.08 & -15.62 & 11:05:37.0& -77:16:38& 4.5 &  0.7 & 0.11&  0.9&   1.7  ~~(0.5)&  0.3\\
 49  & 297.21 & -16.04 & 11:04:27.9& -77:42:41& 4.4 &  0.8 & 0.12&  1.1&   2.5  ~~(0.7)&  0.3\\
 50  & 297.19 & -16.27 & 11:02:16.7& -77:54:18& 4.9 &  0.9 & 0.18&  0.9&   6.0  ~~(1.8)&  0.4\\
 51  & 297.13 & -15.65 & 11:06:05.1& -77:18:58& 4.7 &  0.6 & 0.12&  0.4&   1.2  ~~(0.4)&  0.2\\
 52  & 297.30 & -16.08 & 11:05:43.3& -77:46:36& 3.9 &  0.5 & 0.10&  0.7&   1.1  ~~(0.3)&  0.4\\
 53  & 297.18 & -16.38 & 11:01:11.9& -78:00:11& 5.2 &  0.5 & 0.12&  0.7&   1.4  ~~(0.4)&  0.4\\
 54  & 297.17 & -16.04 & 11:03:48.1& -77:41:10& 4.0 &  0.6 & 0.11&  0.4&   1.0  ~~(0.4)&  0.2\\
 55  & 297.42 & -15.56 & 11:11:27.2& -77:20:45& 4.4 &  0.6 & 0.13&  0.4&   1.6  ~~(0.6)&  0.2\\
 56  & 297.38 & -16.09 & 11:06:54.3& -77:49:15& 4.3 &  0.5 & 0.11&  0.5&   0.8  ~~(0.3)&  0.2\\
 57  & 297.49 & -15.73 & 11:11:25.7& -77:31:51& 4.0 &  0.6 & 0.09&  0.4&   0.7  ~~(0.3)&  0.1\\
 58  & 296.97 & -16.00 & 11:00:53.3& -77:34:05& 4.3 &  0.4 & 0.10&  0.4&   0.6  ~~(0.2)&  0.2\\
 59  & 297.43 & -16.05 & 11:08:02.3& -77:48:05& 4.3 &  0.6 & 0.10&  0.5&   0.9  ~~(0.3)&  0.2\\
 60  & 297.30 & -15.40 & 11:10:42.5& -77:09:09& 3.8 &  0.6 & 0.14&  0.6&   1.6  ~~(0.6)&  0.2\\
 61  & 297.49 & -15.67 & 11:11:46.7& -77:28:15& 4.1 &  0.7 & 0.08&  0.5&   0.8  ~~(0.3)&  0.2\\
 62  & 297.24 & -15.26 & 11:10:35.9& -77:00:28& 3.9 &  0.8 & 0.22&  0.4&   3.8  ~~(1.4)&  0.2\\
 63  & 297.17 & -16.39 & 11:00:52.0& -78:00:35& 4.5 &  0.4 & 0.10&  0.4&   0.7  ~~(0.2)&  0.2\\
 64  & 297.23 & -16.41 & 11:01:46.2& -78:02:48& 4.7 &  0.5 & 0.08&  0.4&   0.6  ~~(0.2)&  0.2\\
 65  & 297.19 & -15.61 & 11:07:27.9& -77:18:22& 3.7 &  0.7 & 0.11&  0.6&   0.5  ~~(0.2)&  0.1\\
 66  & 297.15 & -15.59 & 11:06:51.0& -77:16:17& 4.5 &  0.7 & 0.07&  0.7&   0.7  ~~(0.2)&  0.2\\
 67  & 297.16 & -14.99 & 11:11:21.6& -76:43:23& 4.7 &  0.5 & 0.10&  0.5&   0.7  ~~(0.3)&  0.2\\
 68  & 297.21 & -15.07 & 11:11:35.4& -76:49:07& 4.5 &  0.5 & 0.14&  0.2&   1.4  ~~(0.5)&  0.3\\
 69  & 297.15 & -16.12 & 11:02:51.2& -77:45:21& 4.1 &  0.5 & 0.13&  0.4&   1.2  ~~(0.4)&  0.3\\
 70  & 296.94 & -14.87 & 11:08:43.2& -76:31:35& 4.4 &  0.7 & 0.10&  0.3&   0.8  ~~(0.3)&  0.2\\
 71  & 297.10 & -15.75 & 11:04:52.7& -77:23:48& 3.9 &  0.8 & 0.12&  0.7&   1.0  ~~(0.4)&  0.1
\end{tabular}	    				       			   
}									
\normalsize

\end{table*}
%

%

According to the analysis of \cite{williams1994} the intensity
stepsize, $\Delta T$, for the Clumpfind routine should be set to two
times the maximum noise level. Instead of the absolute maximum, we
used here the 'high' value, 0.15 K, below which the noise level
remains for 90\% of the spectra, and accordingly the intensity stepsize
was set to 0.3 K. This was used also as the intensity threshold for
the clump search and the total cloud mass estimate.

With these parameters the Clumpfind routine identifies 71 clumps in
the dataset. The data cube (``ChaI.fits'') and the clump
identification file (``ChaI.fits.clf''), which is an output of the
Clumpfind programme, will be made electronically  available as FITS files. 
An IDL routine called 'cl\_surf.pro' written by us to
help in inspecting the cloud structure will also be available.
This small programme shows
clumps as isointensity surfaces in the ($x,y,v$)-space using the IDL
Object Graphics and the 'xObjView' interface for zooming and rotating
these surfaces.

The physical characteristics of the clumps  are presented in
Table~\ref{table:clumps}.  The columns of this Table are: (1) the
clump identification number; (2) and (3) the Galactic coordinates of the
clump centre of mass;
(4) and (5) R.A. and Dec. (2000.0); (6) and (7) the LSR velocity of
the line peak and the FWHM of the velocity dispersion in the clump;
(8) the clump half-power radius; (9) the maximum \ceio\ column density; 
(10) the clump mass with 
an error estimate based on the spectral noise; and (11) the
ratio of the gravitational potential to kinetic energy in the
clump. The numbering corresponds to order in decreasing peak
intensity. Clumps 1--6 are detected at the level $7 \times \Delta T$
($14 \sigma$), clumps 7--13 at the level $6 \times \Delta T$, clumps
14--20 at the level $5 \times \Delta T$, clumps 21--36 at the level $4
\times \Delta T$, clumps 37--52 at the level $3 \times \Delta T$, and
clumps 53-71 at the level $2 \times \Delta T$ ($4\sigma$).
  
The division of the data cube into clumps in the central and northern
parts of the cloud is demonstrated in Fig.~\ref{figure:snakes}. 
The clumps identified by the Clumpfind routine are plotted in the figure 
as surfaces in the $(x,y,v)$ space, where $x$ and $y$ are the
R.A. and Dec.  offsets from the $(0,0)$ position, and $v$ is the LSR
velocity.  The total velocity spread in Cha~{\sc i} is only 2.3
$\kmps$ but sometimes two clumps can be separated along the same line
of sight.

The C$^{18}$O column densities (for each $x,y,v$ pixel) were estimated
by assuming optically thin emission, LTE with the excitation
temperature $T_{\rm ex} =10$ K, and a beam-source coupling efficiency,
$\eta_{\rm C}$, of 0.8 (half-way between $\eta_{\rm MB}$ and
$\eta_{\rm Moon}$). These were converted to mass column densities by
assuming that the column density ratio $N({\rm C^{18}O})/N({\rm H}_2)$
is $2\, 10^{-7}$ (\cite{wilson1994}), and that the gas contains 10 \%
He.  The masses were derived by summing up all pixels
within a clump. The distance assumed in the mass calculation is 150
pc. For the gravitational potential energy estimates, it was assumed
that the clump diameter in the radial direction ($z$) is equal to the
smaller of the diameters in $x$ and $y$ directions.  The mutual
gravitational potential energies of all $(x,y,z)$ pixel pairs were
summed up and the internal potentials of pixels (small contribution)
were added by approximating them with homogeneous spheres. The kinetic
energies were also calculated pixel by pixel, and they include the
contributions of systematic and turbulent motions and the internal
thermal energy.

The total mass of the cloud 
(as traced by C$^{18}$O above the threshold 0.30 K) is $230 \,
\Msun$.  The derived clump masses range from 0.5 to 12 $\Msun$ with
the median $2.9 \, \Msun$. The geometric mean radii lie in the range
$0.08 - 0.21$ pc.  The ratios of the gravitational to kinetic energies
are between 0.1 and 1.2, and the median ratio is 0.4.  The masses and
the stability of the clumps will be briefly discussed in
Sect.~\ref{section:discussion}.

The largest uncertainty of the mass estimates is related to the the
$N({\rm C^{18}O})/N({\rm H}_2)$ conversion factor. The value we have
used is close to those derived by \cite{frerking1982} in Taurus and
\cite{harjunpaa1996} in Cha~{\sc i}, and thereby also consistent with
the factor used by \cite{mizuno1999}. The conversion factor is likely
to change towards the centres of dense clumps due to CO depletion. However,
as the column densities are generally modest in this cloud and most
dramatic depletion effects are localized to very high density
regions (e.g. \cite{caselli1999}) we believe that CO
depletion does not cause significant errors to the clump statistics.


\subsection{Star formation and cores}

The locations of known and likely members and candidate
pre-main-sequence stars of the Cha~{\sc i} Association are shown in
Fig.~\ref{figure:stars+c18o}, projected on the C$^{18}$O molecular
line map. The stellar objects were selected from \cite{schwartz1991}
(Optical candidate members, black triangles), \cite{persi2000} (known
members with NIR excess (red squares), member candidates with
NIR excess (open red triangles)), \cite{cambresy1998} (new Young
Stellar Object candidates (asterisks)) and \cite{gomez2001}
(candidate pre-main-sequence stars (open circles)).  Many of the
objects listed in \cite{schwartz1991} were replaced by known members
with NIR excess from \cite{persi2000}. A number of investigations have
been conducted in the direction of the three reflection nebulae known
to be locations of active star formation and more candidate members
could be selected from them.  We think, however, that the objects indicated 
in Fig.~\ref{figure:stars+c18o} reflect well the general distribution of
young stellar objects in Cha~{\sc i}.


   \begin{figure}   \centering   \includegraphics[bb=0 0 600 600,
    width=10cm, angle=0]{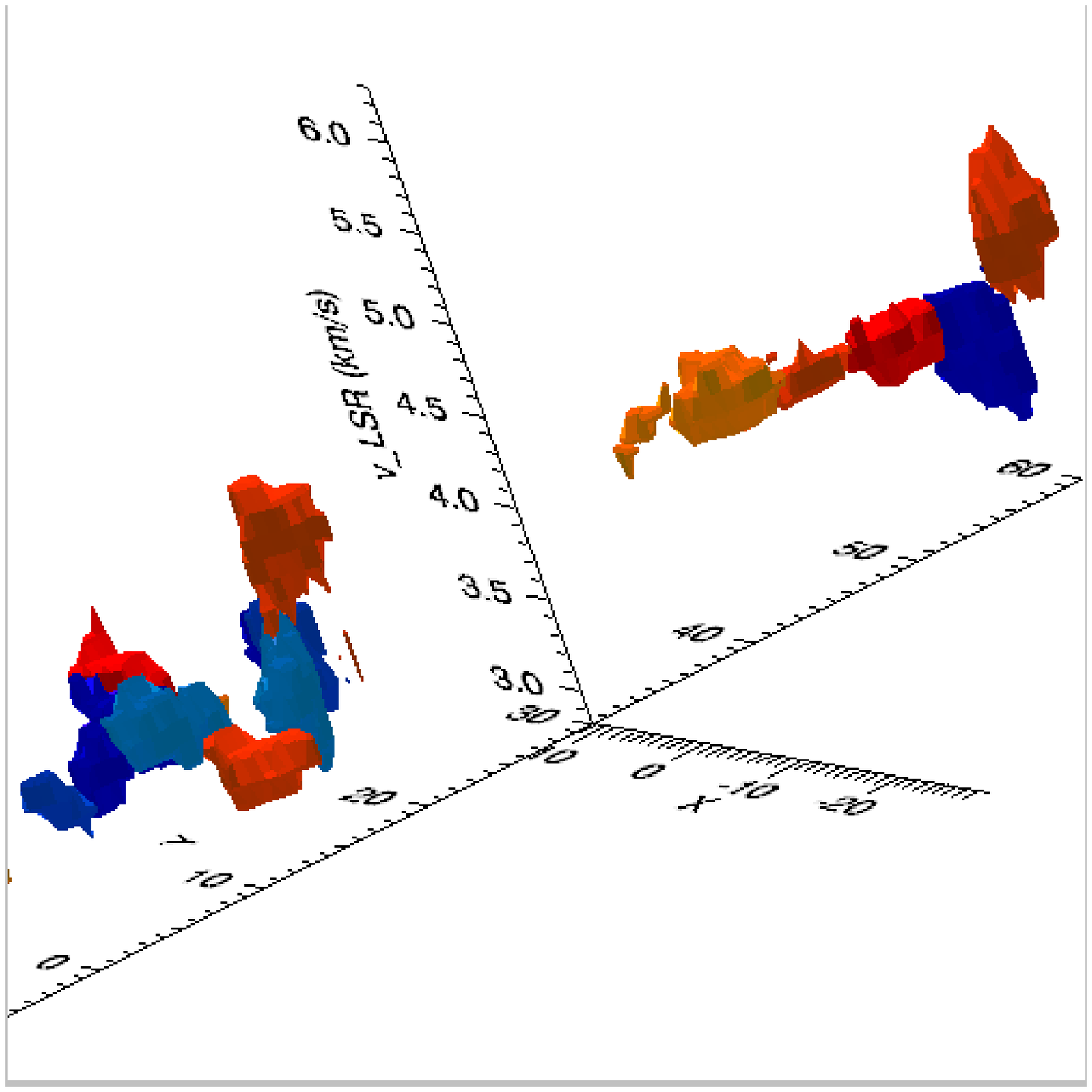}

\caption{Clumps in the central and northern part of Cha~{\sc i}.
          Output of the cl\_surf routine. Each clump identified with
          the Clumpfind programme is shown as an isointensity surface
          in the $(x,y,v)$ space.  Different colours are used to help
          in distinguishing between separate clumps.  The intensity
          level chosen for this figure is $T_A^* = 1$ K. The $x$ and
          $y$ axes represent the offset in arcminutes from the map
          centre (R.A. $= 11\hour 09\minute13\fs3$, 
          Dec. $= -77\degr 29\arcmin 12\arcsec$), and the $v$ axis 
          represents the LSR velocity in $\kmps$. Note the large velocity 
          gradient between the two northernmost clumps 1 and 2 
          at $y \approx 55\arcmin$ (here on the right, the viewing angle
          makes the $y$-offsets look larger) and the 
          exceptionally large velocity of clump 8 in the centre at 
          $y \approx 10\arcmin$  (here on the left).}
\label{figure:snakes} 
\end{figure}
%


   \begin{figure} \centering \includegraphics[bb=80 20 480 750,
   width=9cm]{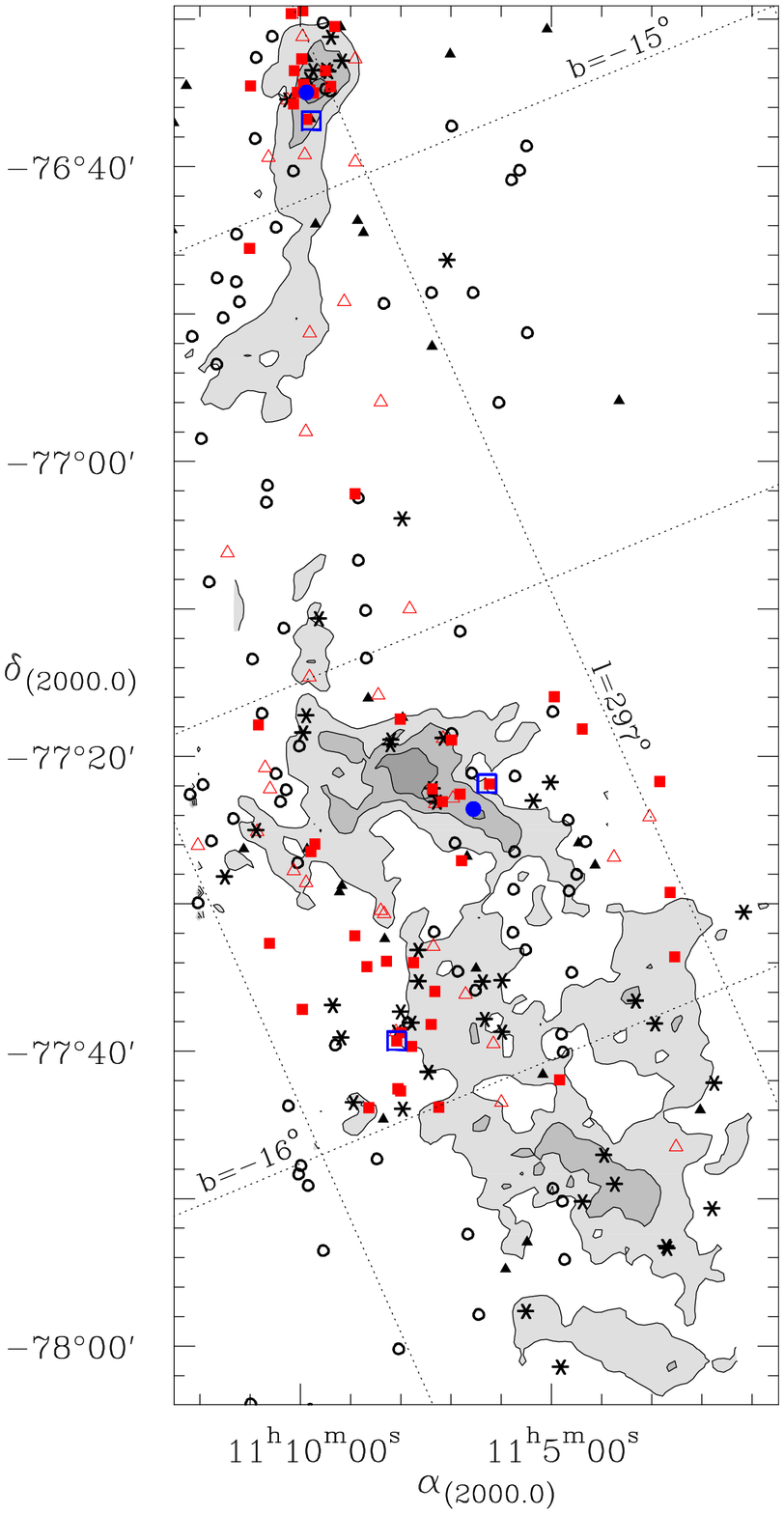} 
   \caption{Cha~{\sc i} association members and
   candidate members projected on the integrated C$^{18}$O intensity
   map of the cloud. Sources of the coordinates are
   \cite{schwartz1991} (Optical candidate members, black  triangles),
   \cite{persi2000} (known members with NIR excess, red
   squares, member candidates with NIR excess, open red triangles),
   \cite{cambresy1998} (new Young Stellar Object candidates,
   asterisks) and \cite{gomez2001} (candidate pre-main-sequence
   stars, open circles).  The locations of the reflection nebulae Ced
   112, Ced 110 and Ced 111 (from north to south) are indicated with
   open squares. The locations of the Class 0 protostar
   candidates Cha-MMS1 and Cha-MMS2 are indicated with filled 
   circles.}
   \label{figure:stars+c18o}
   \end{figure}

Concentrations of young stars are seen in the direction of
the three reflection nebulae, Ced 112 in the north, Ced 110 in the
centre, and Ced 111 near IRN. The stellar clusters near Ced
112 and Ced 110 are associated with massive molecular clumps. Besides
the immediate surroundings of Ced 112 and Ced 110, also the most
prominent molecular filaments in the northern and central parts of the
cloud, i.e. the Northern arm, the $40\arcmin$ Arc, and the central
filament near Ced 110, are lined with young stars. This suggests that
these stellar groupings and large scale molecular structures have a
common origin.

In contrast to the situation near the two other reflection nebulae,
clumps around Ced 111 and the IRN belong to the least massive in the
cloud, and it seems that most of the molecular material around young
stars has dispersed on this side of the cloud.

In the southern parts, south of the galactic latitude $-16\fdg05$, the
surface density of stars is clearly lower than elsewhere in the
cloud. In particular, there is only one ISOCAM mid-IR excess member
candidate source and significantly fewer NIR member candidates than in
the north.  It should be noted, however, that the ISOCAM observations
did not go below $\sim  -77\degr47\arcmin$.  The two parallel southern
filaments were detected in FIR surface emission by \cite{haikala1998}
and \cite{toth2000}.  Considering the small number of NIR and mid-IR
sources, these structures are likely to be cold and their clumps are
possibly in their pre-star formation stage.  (The same might be true
for the starless clump (12) south of Ced 112, even though it lies next to a
string of young stars on its eastern side.)

The large number of newly born stars and their concentration around
filaments seen in the molecular line map indicate efficient
compression of gas and subsequent star formation. According to the
hydrodynamical models (e.q.  \cite{klessen2001}), these features are
characteristic of cloud fragmentation in the conditions of little
turbulent support (decaying turbulence), or turbulence driven by
large-scale shocks. In contrast, turbulent support driven on small
scales leads to inefficient and dispersed birth of stars.
The modest \ceio\ column densities, the fairly large fraction of visible 
stars amongst the associated YSOs, and the fact that only two protostar 
candidates have been found in the cloud, suggest on the other hand 
that we are witnessing the aftermath of star formation. 
The total mass of the young stars associated with the cloud 
has been estimated to be about 120 \Msun\ ({\cite{mizuno1999}), 
which corresponds to about 50\% of the mass of high column density 
gas traced by \ceio.

\subsection{Orientation of the magnetic fields}

%

%
   \begin{figure} \centering \includegraphics[bb=50 0 670 810,
   width=12cm]{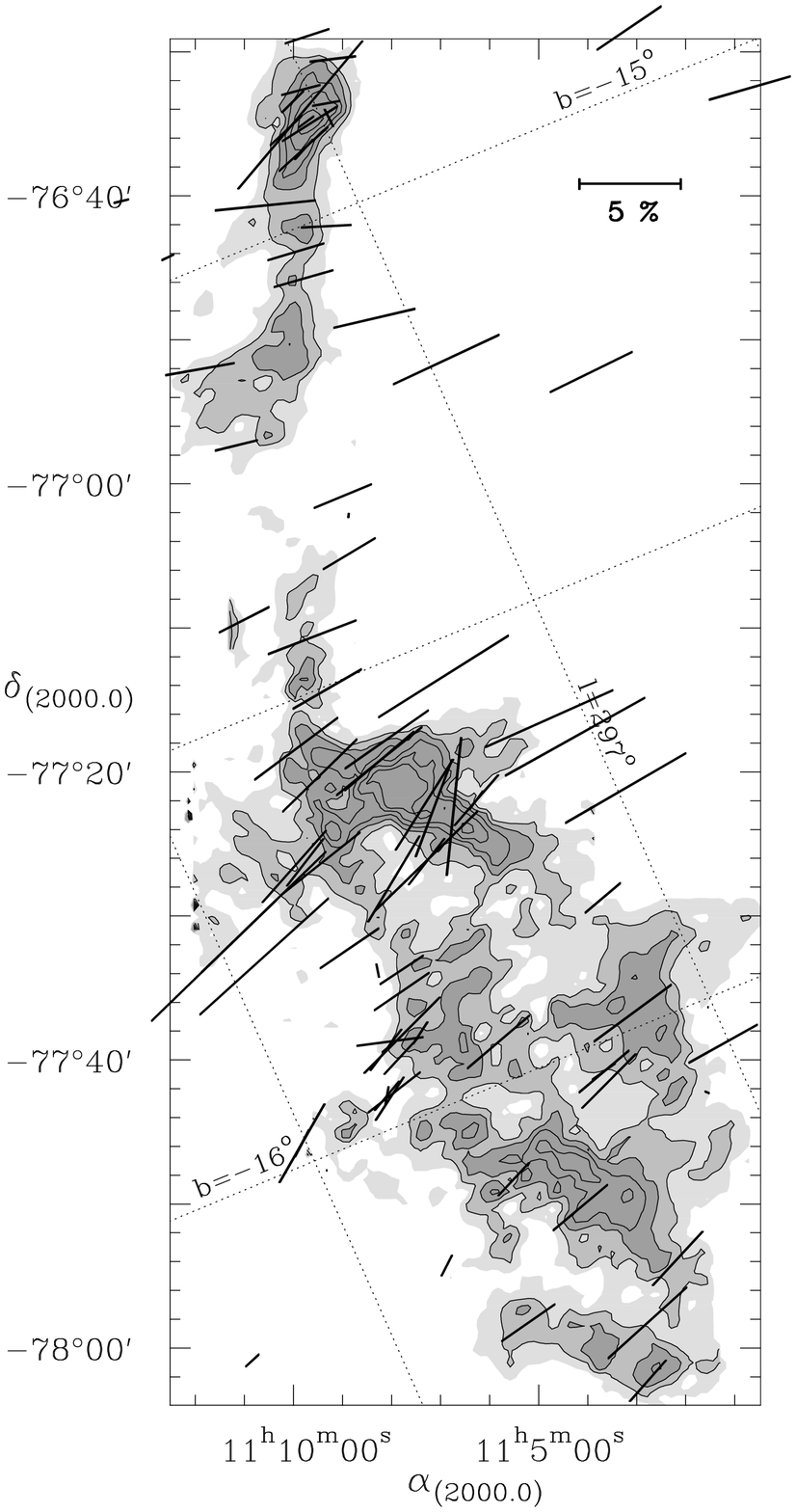} 
   \caption{Polarization vectors of  field  stars and  Cha~{\sc I}
   members (\cite{whittet1994}, \cite{mcgregor1994})
   superposed on the integrated C$^{18}$O intensity map of Cha~{\sc
   i}.}
   \label{figure:pol+c18o}
   \end{figure}
%

Near-infrared ($H$-band)
polarimetry of heavily reddened background stars and embedded
objects was  used by \cite{mcgregor1994} to study the magnetic field 
structure in Cha~{\sc i}.
The stellar polarization 
due to aligned dust particles in the direction of the cloud is 
illustrated in Fig.~\ref{figure:pol+c18o}. The data is from 
\cite{mcgregor1994} and \cite{whittet1994}.
 The polarizations of  some of  the Cha~{\sc I} members are 
influenced by intrinsic polarization due to scattering in the circumstellar
dust or the surface of the molecular cloud. Depending on the geometry, this
can lead to depolarization and/or rotation of the polarization direction.
A notable example is the  IRN (polarization not  plotted in Fig.~ 
\ref{figure:pol+c18o}) for which the $H$ band polarization 
is 33\% and the polarization angle differs by tens of degrees from the
cloud average (\cite{mcgregor1994}).
The small foreground  reddening in the direction of Cha~{\sc i}
and the high galactic latitude makes it certain that the
observed high polarizations are in major part due to dust in the Chamaeleon 
region. These observations probe magnetic fields in the obscured  parts
of the cloud and have thus a direct bearing on structures revealed by
the present \ceio\ mapping. The polarization  field is well ordered
throughout the cloud, changing its orientation from nearly E-W in the
northern part to NW-SE in the centre and in the south. This
orientation, which is consistent with the large-scale polarization field
in the region (\cite{whittet1994}), is roughly perpendicular to the cloud 
axis, the dense filaments seen in \ceio \  and presumably also to the magnetic 
field in the region.  
As pointed out by \cite{mcgregor1994}, the assumed  orientation of the magnetic
field suggests that it has steered the cloud collapse preferentially
along the field lines. That also dense filaments seem to be
oriented at right angles to this  magnetic field lends support to this
view.

It should be noted, furthermore, that turbulent fragmentation models
predict that the resulting magnetic field in dense filaments or sheets
is parallel to the direction of elongation, because this magnetic
field component is amplified in shocks (\cite{padoan2002}). The
$H$-polarization percentage versus visual extinction in Cha~{\sc i}
seems to saturate at $A_V \sim 7$ (see Fig.~4 of \cite{mcgregor1994}),
suggesting that the grain alignment weakens in the densest parts
(see the discussion in \cite{goodman1992}).

\section{Comparison with NANTEN  cores}

The integrated C$^{18}$O$(J=1-0)$ intensity map presented in
Fig.~\ref{figure:chai_c18o} shows the same general outline as the one
taken with the 4-m NANTEN telescope (\cite{mizuno1999}).  As expected,
the higher spatial resolution of the SEST allows one to resolve more
details.  The cores detected by \cite{mizuno1999} are all seen in the
SEST data but these cores are fragmented into smaller entities, both
spatially and in velocity.

The correspondence between 'cores' identified by \cite{mizuno1999} in
the NANTEN map, and 'clumps' identified with the Clumpfind programme
in the SEST map is most clear in the northern arm or Cha~{\sc i}
North.  Our clumps 1 and 2 in Cha~{\sc i} North with a total mass of
about $10\,\Msun$ correspond to the centre of NANTEN core 3 (N3), for
which they derived the mass $29\,\Msun$.  The velocity difference
between clumps 1 and 2 is clearly visible in Fig.~\ref{figure:snakes}
(at $y \approx 55\arcmin$).  Clump 12 ($12\,\Msun$) about $15\arcmin$
south of Cha~{\sc i} North corresponds to core N4 ($16\,\Msun$).  The
SEST map shows, however, several subsidiary peaks also in the north,
and especially the windswept shape of the northern filament is 
clearly visible there.

In Cha~{\sc i} Centre and in the southern part of the cloud, where the
structure is more complex, the two NANTEN cores (6 and 5) decompose
into several clumps.  Towards the absolute intensity maximum of the
cloud, the peak of core N6, two clumps lie in the line of
sight at different radial velocities (clumps 8 and 17 at 
$y \approx 10\arcmin$ in Fig.~\ref{figure:snakes}). The existence
of two velocity components was already noted by
\cite{mizuno1999}. Besides these two, core N6 comprises two
major clumps (3, 19) and has several smaller clumps on
its western side. The total mass of the four most massive clumps is 
$30\,\Msun$, which roughly corresponds to the mass derived for Mizuno's
core 6, $39\,\Msun$. 

The centre of core N5 in the south correspond to our clumps 5 and 6
which belong to one of the two parallel filaments in the south
\footnote{Note: The coordinates of clump 5
in Table~1. of \cite{mizuno1999} are not correct. 
The correct coordinates are  $l,b = 297\fdg23, -16\fdg13$ (Mizuno 2003, private
communication.)}. 
The procedure we have used identifies half a dozen of clumps in this filament 
with a total mass of $30\,\Msun$, which is well below the mass derived for 
core N5
($82\,\Msun$) by \cite{mizuno1999}. The most massive individual clumps in
the two southern filaments are numbers 6 and 16,
respectively, which also are among the most massive in the whole
cloud. 

\section{Notes on individual regions}

\subsection{Cha~{\sc i} North}

The Clumpfind program finds in this region three major clumps, 1, 2,
10.  Clump 1 is in the direction of the B9V star HD97300, which is the
illuminating source of Ced 112. Clump 2 corresponds to the opaque core
described in \cite{jones1985}. It contains the Class 0 protostar
candidate Cha-MMS2 (\cite{reipurth1996}), which is a possible driving
source of the molecular outflow detected in this region
(\cite{mattila1989}). Cha-MMS2 lies in the apex of the blueshifted
outflow lobe.  Numerous mid-IR sources with mid-IR excess are seen
projected in the line of sight to clumps 1 and 2
(\cite{persi1999}; \cite{persi2000}). One of these sources,
ISOCAM-ChaINa2 has a SED characteristic of a Class I source and has
been proposed by \cite{persi1999} to be the possible exciting source of
the molecular outflow.

Clump 12 (the centre of N4), which lies $15^{\prime}$
south of clumps 1 and 2 is devoid of mid-IR sources and does not seem
to be connected with star forming activity. This core was detected
also in the ISOPHOT 170 $\mu$m serendipity survey of the Chamaeleon
clouds (\cite{toth2000}), which indicates the presence of very cold
dust in this clump.

\subsection{Cha~{\sc i} Centre}

The prominent intensity maximum near Ced 110 (N6) is split into
four major clumps: 3, 8, 17 and 19.  Projected on the sky, clump 8
partially overlaps with clump 17, but it has a larger radial velocity
(see Fig~\ref{figure:snakes}). This region contains a cluster of
low-mass young stellar objects which have been studied extensively in
recent years (e.g. \cite{prusti1991}; \cite{persi2000};
\cite{persi2001}; \cite{lehtinen2001}; \cite{lehtinen2003}). The dense
dust ridge detected at 200 $\mu$m with ISOPHOT (\cite{lehtinen2001})
coincides with the central parts of clumps 3 and 8. The Class 0
candidate Cha-MMS1 (\cite{reipurth1996}) is probably embedded in clump
3, close to its north-eastern boundary.  \cite{reipurth1996} suggested
that Cha-MMS1 is the central source of the bipolar molecular outflow
discovered by \cite{mattila1989}, whereas \cite{lehtinen2003} regarded
the Class I infrared source IRS~4 as a more likely candidate. In the
latter case the outflow would originate between clumps 3 and 17.

The clumps 60, 35, 19, 33, 18, 27 and 25, together
with  the clumps west of the Ced 111 region, 22,
23 and 30, form a nearly continuous arc of 40$\arcmin$. The
clumps in the arc are however readily separated both spatially and in
velocity (Fig.\ \ref{figure:chai_c18o_channel}).

\subsection{Ced 111 region}

Besides the Ced 111 reflection nebula, the southeastern edge of the
cloud contains the IRN and several pre-main-sequence stars
(eg. \cite{schwartz1991}). Only four clumps, 22, 23, 30 and 38 are located in
the region which has formerly been a centre of active star
formation. It seems that this process has consumed and dispersed most
of dense material of the ambient cloud. The clump 38, which is detected at a
low level, is associated with the IRN.

\subsection{Cha~{\sc i} West and South}

 The clumps 4, 7, 9, 13, 14 and 15 to the West of Ced 111 form
a massive condensation of clumps. This conglomeration of clumps is
probably  associated with the Very cold Core 5 (VCC5) in \cite{toth2000}.
The two parallel elongated structures to the the South 
are clearly  seen in the far-IR (\cite{haikala1998}, \cite{toth2000}).
\cite{toth2000} designated these structures as  VCC4 and VCC3. 
Similar to  VCC4 also the two other separate into
smaller  units. The clumps 5 and 6 correspond to VCC4 and clumps
16 and  24 to VCC3. 
None of these structures contain IRAS point sources with colours typical 
of newly born stars, and seem to have no active star formation.  
The ISOCAM mid-IR
Chamaeleon mapping (\cite{persi2000}) covered only  the northern part of
these regions.

\section{Distribution of clump masses and their stability}
\label{section:discussion}

\begin{figure}
\centering
\includegraphics[bb=120 350 570 720,width=11cm]{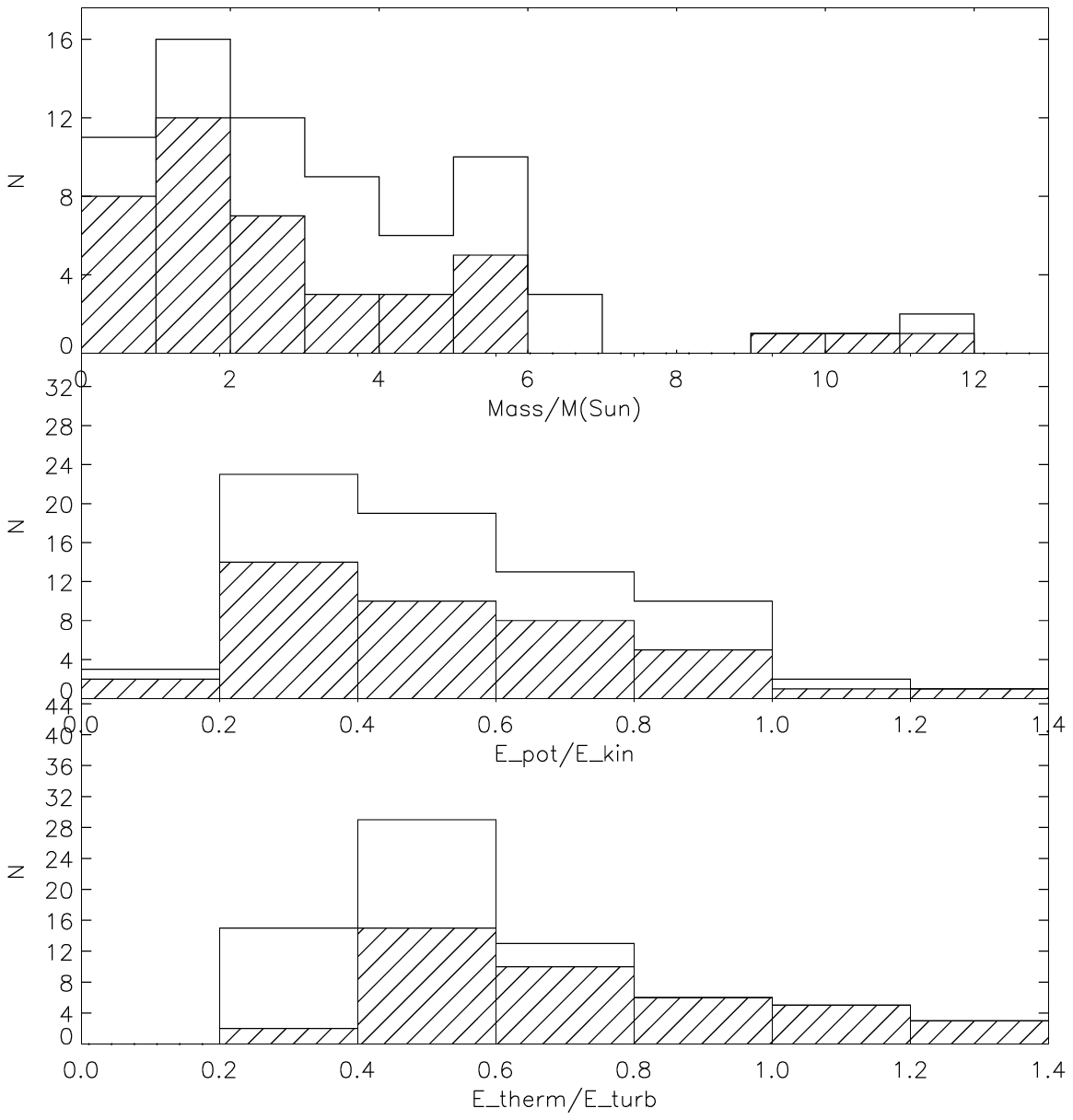}
\caption{{\bf Top:} Distribution of the clump masses in
Cha~{\sc i}.  {\bf Middle:} The ratios of the gravitational
potential to the kinetic energies, $|E_{\rm pot}|/E_{\rm kin}|$. The
kinetic energy estimates contain both the internal thermal energy
and the energy of the turbulent motions.  {\bf Bottom:} The
ratios of thermal to turbulent energies.  The hatched columns
correspond to clumps which possibly are in virial equilibrium (see text).}
\label{figure:cha_histograms}
\end{figure}

 The distributions of the derived clump masses, the ratios of the
potential to kinetic energies, and the ratios of the thermal to
turbulent kinetic energies are shown in
Fig.~\ref{figure:cha_histograms}.  The clump masses range from 0.6 to
12 $\Msun$ with the median $2.7 \, \Msun$.  The $|E_{\rm pot}|/E_{\rm
kin}$ ratios are between 0.1 and 1.2, and the median ratio is 0.4.
The ratio is close to unity for the most massive clumps.  For most
clumps the turbulent energy dominates over the thermal energy.

As the condition of the virial equilibrium (in the absence of an
external pressure) is $|E_{\rm pot}|/E_{\rm kin} = 2$, the result
indicates that, strictly speaking, none of the clumps identified
here is a gravitationally bound entity. They are either dissolving or
stabilized by the outside pressure due to interclump turbulence or
by magnetic fields.

\begin{figure}
\centering
\includegraphics[bb=100 370 550 700,width=9cm]{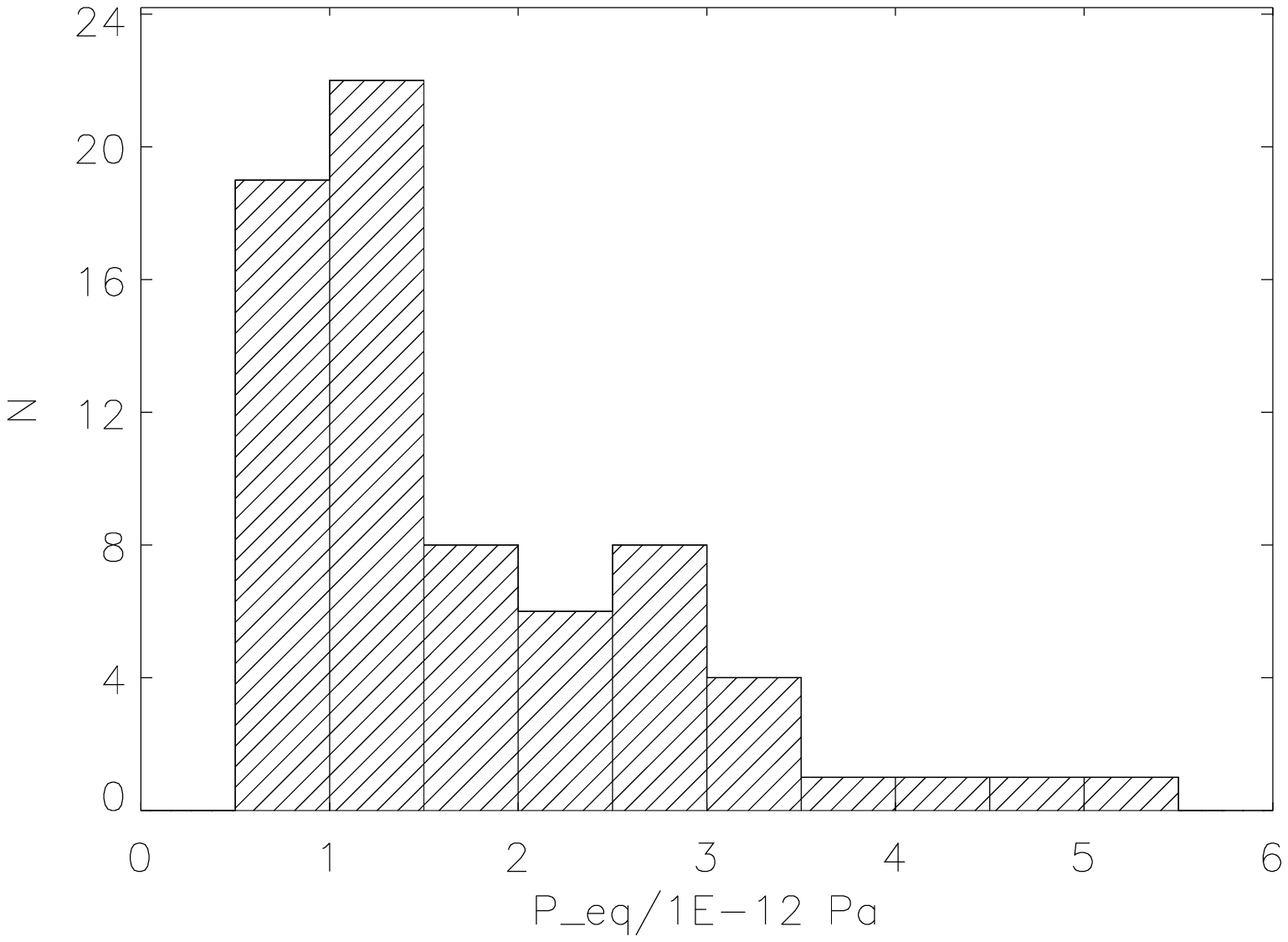}
\caption{The distribution of external pressures, $P_{\rm ext}$ 
needed to stabilize the identified clumps.}
\label{figure:peq_histo}
\end{figure}

We may estimate for each clump a hypothetical external pressure,
$P_{\rm ext}$, required to balance the 'extra' turbulent motions using
the virial equilibrium equation:
\begin{equation}
2 E_{\rm kin} + E_{\rm pot} = 4\pi R^3 \, P_{\rm ext} \; . 
\label{eq:virial}
\end{equation}

The distribution of these 'equilibrium pressures' is shown in
Fig.~\ref{figure:peq_histo}. The distribution is asymmetric, peaking
towards a minimum which lies close to the value $10^{-12}$ Pa ($P/k
\sim 7\,10^4$ cm$^{-3}$K).  
From observational statistics it is likely that the external pressure
may be sufficient to stabilize some of the clumps, but not to force
any of them to collapse.
In case the pressure of the interclump medium is roughly
constant across the cloud, the minimum of the derived $P_{\rm ext}$
values should correspond to the actual interclump pressure. 
One can see in Fig.~\ref{figure:peq_histo} that
in fact for about 40 clumps $P_{\rm ext}$ is close to $10^{-12}$ Pa. 
 We assume therefore that this value represents the typical interclump
pressure, and that those 40 clumps are in the virial equilibrium. The mass
distribution of these 40 clumps is presented as hatched columns in
Fig.~\ref{figure:cha_histograms}. It is similar to that of the
original sample, but these clumps have less marked turbulent motions
(see the bottom panel of Fig.~\ref{figure:cha_histograms}).
In case the clumps are stabilized by the magnetic field,
the field strength is 16 $\mu$G, as can be obtained from the
following formula for the magnetic pressure:
$P_{\rm m} = B^2[\mu \rm G]/(8\pi) \, 10^{-13}$ Pa.

The clump mass spectrum can be, in principle, used for studying the
mechanisms of cloud fragmentation, and, if sufficiently small scale
structures can be resolved, also for predicting the stellar initial
mass function, IMF (see e.g. \cite{padoan2002};
\cite{ballesteros2002}; \cite{klessen2001}).  The spatial resolution
and the RMS noise level of the present data are clearly not sufficient
for studying the complete mass spectrum down to clumps at $\leq 0.5
\Msun$.  Stars may form in the interiors of the larger,
quasi-stable clumps identified here, but the relation between their
masses and the stellar mass function is beyond the scope of this
paper.

We discuss briefly, however, the possibility to understand the mass
spectrum of the hypothetically virialized clumps in terms of the
turbulent fragmentation model presented in \cite{padoan1997} and
\cite{padoan2002}. These studies predict 1) the mass distribution of
dense cores, 2) the probability density function of the gas density, and
3) the subsequent mass distribution of collapsing clumps arising from
supersonic turbulence.

In the model of \cite{padoan2002} the mass distribution of {\sl all}
clumps follows a power law depending on the power spectrum of the
turbulence: $N(m) \propto m^{-3/(4-\beta)}$, where $\beta$ is the
spectral index of the turbulence (see their Eqs. (5) and (18)).  Using
their assumption that the density and mass distributions are
statistically independent, one can derive the distributions of the
Jeans' masses, $\Phi_{\rm J}(m)$, or alternatively, the distribution of
'equilibrium masses' $\Phi_{\rm E}(m)$, under the condition of
constant external pressure expressed in Eq.~\ref{eq:virial}.  In the
latter case the equivalent of Eq. (24) of \cite{padoan2002} can be
written as:

\begin{equation}
\Phi_{\rm E}(m) \propto m^{-3/(4-\beta)} \; 
\int_{\tilde{m}=0}^m \, g(\tilde{m}) \, d \tilde{m} \; ,
\label{eq:fim}
\end{equation}
where $g(\tilde{m})$ is the probability density function of
equilibrium masses, and $\tilde{m}$ is the variable of
integration. The function $g$ is derived below.  The observed
distribution of pressure balanced clumps (in the top of
Fig.~\ref{figure:cha_histograms}) should be a sample of the
distribution $\Phi_{\rm E}$.

The $\beta$-index should be reflected in the observed size-linewidth
relation, $\Delta v \propto R^\alpha$, since $\alpha=(\beta-1)/2$
according to Eq.(13) of \cite{padoan2002}. The expected values range
from 1.6 (for incompressible turbulence) to 2 (shock dominated
turbulence; \cite{larson1979}; \cite{larson1981}).  The clumps
identified in the present study show a very weak dependence between
the line width and size, and a large scatter.  A least-squares fit to
the identified clumps gives $\alpha = 0.16 \pm 0.09$ (instead the
usual $\alpha \approx 0.5$) with a small correlation coefficient of
$r=0.2$. The AOS channel width and the mapping step size
correspond to 0.12 \kmps \ and 0.044 pc, respectively. These values
correspond to a large fraction of the total range of the observed line
widths (from 0.4 to 0.9 \kmps) and sizes (from 0.08 to 0.22
pc). Therefore, even if a correlation between the line widths and
clump sizes exists in the cloud, the present data are not suitable to
make a meaningful fit when the errors are taken into account.  The
qualitative results presented here do not depend, however, strongly on
the actual value of $\beta$. Since its determination from observations
is furthermore subject to large uncertainties, we assume in the
following the value 2 valid for compressible turbulence.

We next derive the function $g(m)$ appearing on the
right hand side of Eq.~\ref{eq:fim}.  According to the turbulent
fragmentation models of Padoan et al., the probability density
distribution of the local density parameter, $p(x)$, where $x\equiv
n/\bar{n}$, and $\bar{n}$ is the average number density in the cloud,
follows approximately a lognormal function. The probability density
distribution {\sl per unit mass}, i.e. the mass function at any given
density, $f(x) = x p(x)$, can then be written as
\begin{equation}
f(x) = \frac{1}{\sqrt{2\pi}\sigma_{\ln x}} \, 
\exp \{-\frac{1}{2} \left(\frac{\ln x - \overline{\ln x}}
{\sigma_{\ln x}}\right)^2 
\} \; .
\label{eq:fx}
\end{equation}
(see Eqs. (1) and (7) of \cite{padoan1997}). As pointed out in
the latter study, the properties of the lognormal function and the
selection of the variable $x$ such that $\bar{x} =1$ imply that the
mean, $\overline{\ln x}$, and the standard deviation, $\sigma_{\ln x}$, are
related by $\overline{\ln x}=-\sigma_{\ln x}^2/2$.  The distribution is thus
characterized by a single parameter, $\sigma_{\ln x}$. This in turn
depends on the rms Mach number, ${\cal M}$, of the turbulent flow 
according to $\sigma_{\ln x}^2 \approx \ln (1 + 0.25 {\cal M}^2)$
(\cite{padoan1997}, Eq. (9)), i.e. inhomogeneity grows with 
the speed of the flow. 

\begin{figure*}
\centering
\includegraphics[bb=60 360 540 700,width=17cm]{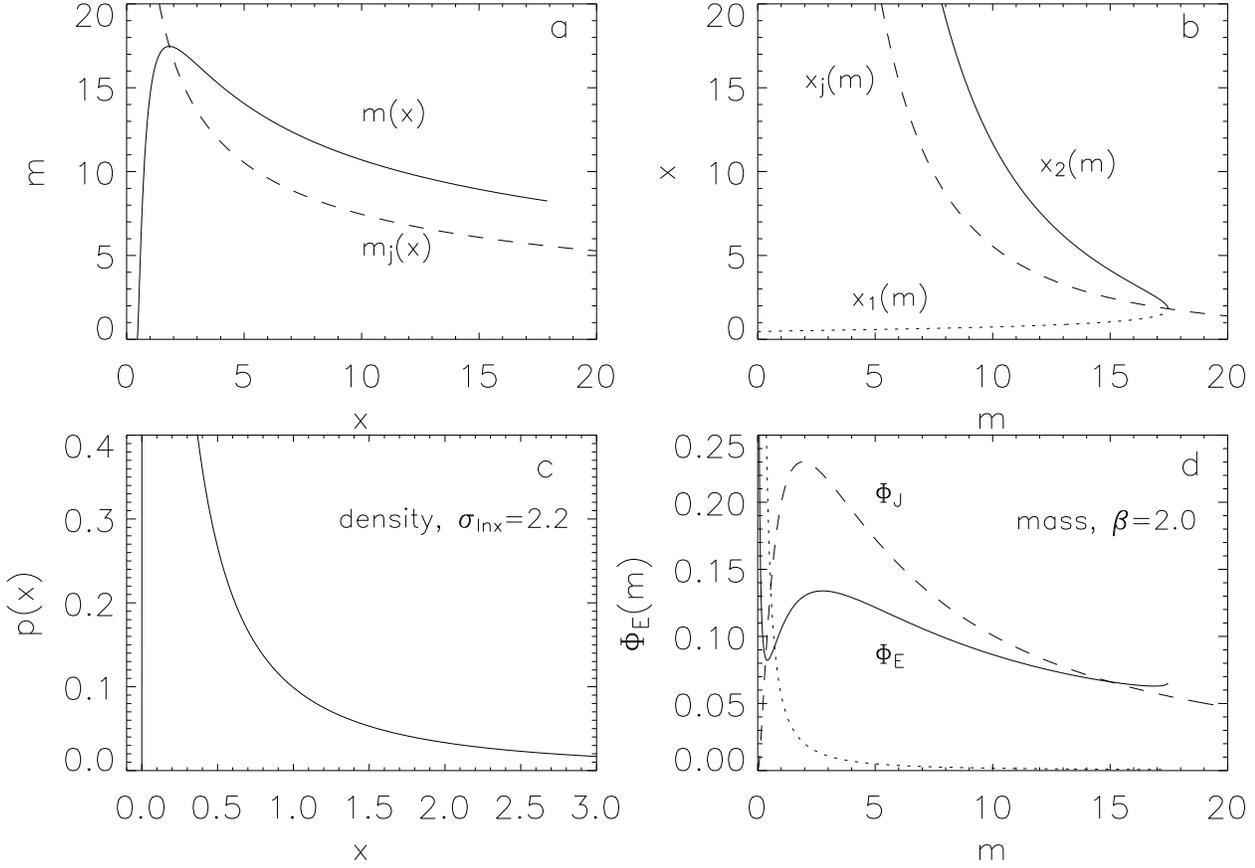}
\caption{{\bf a)} The 'equilibrium mass' $m(x)$, i.e. the clump
mass as a function of the density parameter $x\equiv n/\bar{n}$ (solid
line). The physical parameters used here correspond to values estimated 
for the \ceio\ clumps in Cha~{\sc i}: $P_{\rm ext}= 1 \, 10^{-12}$
Pa, $\bar{n} = 5.5 \, 10^{3} \; {\rm cm}^{-3}$, $T = 10$ K, $\sigma_v =
260 \; {\rm ms}^{-1}$. The Jeans' mass, $m_{\rm j}$, for these 
conditions is plotted with a dashed line. ~{\bf b)}
Density parameter $x$ as a function of mass, i.e. the inverse of
$m(x)$ in two regions $x_1(m)$ (dotted line, low densities) and
$x_2(m)$ (solid line,high densities). The corresponding function for
Jeans' masses, $x_{\rm j}$, is plotted with the dashed line.  ~{\bf
c)} The assumed lognormal density distribution, $p(x)$,
with the standard deviation $\sigma_{\ln x} = 2.2$.  ~{\bf d)} 
The probability density function of equilibrium masses,
$\Phi_{\rm E}(m)$. The distribution of the Jeans' masses, 
$\Phi_{\rm J}(m)$, is plotted as a dashed curve. The dotted curve 
represents the assumed mass distribution of all dense cores (not in scale). 
It follows the power law $m^{-1.5}$, which is consistent with compressing
turbulence.}
\label{figure:mx+xm} 
\end{figure*}

Turning to a small element of the cloud, we examine 
an isothermal, homogeneous clump formed within it. 
From Eq.~\ref {eq:virial} one can derive the following expression
for the mass of a spherical, virialized clump as a function of the 
density parameter $x$:

\begin{equation}
m(x) = a x^{-1/2} (b-cx^{-1})^{3/2} \; ,
\label{eq:mx}
\end{equation}
where 
\begin{eqnarray}
a & \equiv &  \left(\frac{5}{G}\right)^{3/2} 
              \left(\frac{3}{4\pi \mu \bar{n}}\right)^{1/2} \nonumber \\
b & \equiv &  \frac{k T}{\mu} + \sigma_v^2  \label{eq:constants} \\
c & \equiv &  \frac{P_{\rm ext}}{\mu \bar{n}} \nonumber  \; , 
\end{eqnarray}
where $G$ is the gravitational constant, $k$ is the Boltzmann
constant, $T$ is the gas kinetic temperature, $\mu$ is the mean
molecular mass, and $\sigma_v$ is the turbulent velocity dispersion
inside the clump. For Jeans' masses the corresponding function can be
written as $m_j(x) = (\frac{3}{4})^{3/2} a b^{3/2} x^{-1/2}$ with the
same definitions of $a$ and $b$ as in Eq.~\ref{eq:constants}.  One can see from
Eq.~\ref {eq:mx} that a minimum density, $x_0 = c/b$, is required for
an equilibrium mass. This condition is equivalent with the requirement
that the internal pressure, $P_{\rm int} \equiv n \mu (kT/\mu +
\sigma_v^2)$, must be equal to or larger than $P_{\rm ext}$.  The
equilibrium mass reaches a maximum at $x=4 x_0 = 4c/b$, where 
$m = (3^{3/2}/16) \, a \, b^2 \, c^{-1/2}$, and decreases monotonically
thereafter.  At the maximum $m(x) = m_j(x)$ and $E_{\rm pot} =
-\frac{3}{2} E_{\rm kin}$. This turning point signifies the density
at which gravity starts to dominate over external pressure as the 
counterforce against the internal pressure. 

The functions $m(x)$ and $m_j(x)$ are plotted in
Fig.~\ref{figure:mx+xm}a.  The kinetic temperature is assumed to be 10
K, and the average gas density and velocity dispersion derived for the
\ceio\ clumps ($\bar{n} = 5500$ \percc, $\sigma_v = 260$ ms$^-1$) are
used to estimate the values of the parameters $a$, $b$ and $c$.  It
can be seen from this figure that $m(x)$ exceeds $m_j(x)$ when 
$x > 4c/b$. According to the definition of Jeans' mass this
means that the equilibrium masses on the right hand side of the
maximum will collapse if the external pressure increases.

The probability density function of equilibrium masses, $g(m)$, can be
derived from Eqs.~\ref{eq:fx} and \ref{eq:mx}. For each mass below the
maximum derived above there are two possible values of the density
parameter $x$, say $x_1$ (low density, $< 4 x_0$) and $x_2$ (high
density, $> 4 x_0$, $4 x_0$ corresponds to about $10^{4}$ \percc\ with
the parameters used for Fig.~\ref{figure:mx+xm}).

If we denote by $M$ and $X$ the random variables
describing the mass and the density parameter, respectively, the
probability distribution function of the mass can be expressed with
the aid of the density as follows:
\begin{eqnarray}
P(M \leq m) &=& P(X \leq x_1) + P(X \geq x_2) \nonumber \\
            &=& \int_0^{x_1(m)} f(x) dx +  \int_{x_2(m)}^\infty f(x) dx \; .
\label{eq:PM}
\end{eqnarray}  
The probability density function, $g(m)$ can  
then be obtained from the latter by derivation:
\begin{eqnarray}
g(m) &=& f(x_1(m)) \frac{d x_1(m)}{d m} - 
         f(x_2(m)) \frac{d x_2(m)}{d m} 
 \nonumber \\
     &=& \frac{f(x_1(m))}{m^\prime(x_1(m))} - \frac{f(x_2(m))}
         {m^\prime(x_2(m))} \; ,
\label{eq:gm}
\end{eqnarray}
where function $f$ is defined by Eq.~\ref{eq:fx}, $x_1(m)$ and
$x_2(m)$ are the inverse functions of $m(x)$ in the regions $x_0 \leq
x \leq 4 x_0$ (low densities) and $x > 4 x_0$ (high densities),
respectively, and $m^\prime$ is the derivative of function $m$.  The
expressions of $x_1(m)$ and $x_2(m)$ that follow directly from
Eq.~\ref{eq:mx} are rather intricate and they are not shown here. The
functions $x_1(m)$ and $x_2(m)$ are, however, plotted in
Fig.~\ref{figure:mx+xm}b, together with the inverse of Jeans' mass,
$x_j(m)$.

Estimates for the parameters $a$, $b$, $c$ and $\beta$ in
Eqs.~\ref{eq:mx} and \ref{eq:fim} can be derived from observations or,
as in the case of $\beta$, from theoretical predictions.  The mass
distributions of equilibrium and collapsing clumps predicted by the
model of Padoan et al. depend then ultimately on the gas density
distribution described by the parameter $\sigma_{\ln x}$.  It turns
out that a small dispersion in densities ($\sigma_{\ln x} < 1.2$)
makes the Jeans' mass distribution peak towards high masses, $M>10
\Msun$, which contradicts the observed stellar mass distribution in
Cha~{\sc i} and in other dark clouds.  A situation where both Jeans'
masses and equilibrium clump masses peak below $1\, \Msun$ require a
wide range of densities, and consequently, a large value of
$\sigma_{\ln x}$, typically in excess of 2.5. These high values in
turn imply large turbulent Mach numbers (${\cal M} > 45$,
corresponding to 8.6 \kmps\ at 10 K or 30 \kmps at 100 K;
\cite{padoan1997}, Eqs. 4 and 5).  This situation may have resulted
from the passage of a powerful shock wave, e.g. one caused by a
collision with an expanding bubble driven by supernova explosions and
stellar winds from an OB association.

The density distribution for a slightly smaller density dispersion,
$\sigma_{\ln x}=2.2$, and the resulting mass distribution are
illustrated in Figs.~\ref{figure:mx+xm}c and d, respectively. The
latter Figure shows the probability density function of 'equilibrium
masses', $\Phi_{\rm E}(m)$, obtained from Eqs.~\ref{eq:fim} and
\ref{eq:gm}, and the corresponding distribution of Jeans' masses,
$\Phi_{\rm J}(m)$,w together with the 'mother' distribution of all
clumps, assumed to follow the power law $m^{-1.5}$.  The value of
$\sigma_{\ln x}$ is chosen so that $\Phi_{\rm E}(m)$ decreases towards
higher masses, but the function still shows the characteristic peaking
towards zero, which cannot be distinguished at larger values of
$\sigma_{\ln x}$.  The peak near zero is the contribution of
low-density, pressure balanced clumps, whereas the smooth curve
peaking near $3 \Msun$ represents dense, gravitationally bound
clumps. The latter distribution resembles that of the Jeans' masses,
but is flatter.

The clumps detected in this study are likely to bath in 'interclump'
gas with the density just below the critical density of the
C$^{18}$O($J=1-0$) line, i.e. about $10^3$ cm$^{-3}$.  Assuming that
the external pressure $10^{-12}$ Pa is caused by turbulence in this
interclump gas, and that the gas temperature remains at 10 K also
there, we find that the required turbulent velocity dispersion is 0.5
\kmps\ or that the rms Mach number is about 3, i.e. 15 times lower
than the one consistent with the observed {\sl mass} distribution
according to the model of \cite{padoan2002}.

The discrepancy may be partly explained by the hierarchical structure
of turbulence. If the rms Mach number 3 is characteristic for the
length scale 0.1 pc (corresponding to the typical size of a clump),
${\cal M} = 45$ ($\sigma_v \approx 8.5$ \kmps) should then
characterize turbulence in the length scale 20 pc, assuming that the
scaling law is $\sigma_v \propto L^{0.5}$ valid for compressible
turbulence.  This length scale corresponds to the size of the whole
Chamaeleon dark cloud complex including the clouds {\sc i}, {\sc ii}
and {\sc iii}.  The time scale, $L/\sigma$, associated with it is 3
Myr, meaning that in this time the large-scale turbulence should have
decayed into motions on small scales. The ages of the star
associations do not contradict with this time scale.  As discussed in
Sect. 3.2., a large fraction of the stars associated with the Cha~{\sc
i} are visible pre-main-sequence stars, which indicates that at least
a time on the order of $10^6$ years has passed since the initiation of
star formation and the possible violent event triggering it. 
On the other hand, if the quoted scaling law is valid, one could expect to
see a larger dispersion in the radial velocities of the clumps across
the cloud, i.e. over a projected distance of about 4 pc, than the value 
0.35 \kmps\ mentioned earlier in this Section.

\vspace{2mm}
 
The exercise performed above shows that the origin of the observed
clump mass distribution may be explained by the turbulent
fragmentation model of \cite{padoan2002} if turbulence has initially
been hypersonic, but cascaded down to much smaller scales and speeds
by now. The deduced present-day velocity dispersion in the interclump
gas suggest that the length scale has initially been larger than the
whole Cha~{\sc i} cloud.  According to numerical simulations
turbulence acting on large scales gives rise to extensive filaments of
dense gas (e.g.  \cite{klessen2003}), and the structure observed in
Cha~{\sc i} could be readily understood. However, as discussed in
Sect. 3.3 the well-ordered magnetic field is not consistent with this
picture. Moreover, it is not clear from simulations if the velocity
dispersion across the cloud can become as low as observed in Cha~{\sc
i} in the turbulent fragmentation process.  

In view of the fact that the region studied here may cover only a
small fraction of the structure formed in this process, it is
questionable how well it can reflect the global properties of the
flow.  We therefore do not proceed to a quantitative analysis, but
intend to present such in a subsequent paper with a more varied sample
of clumps, obtained by combining results from different clouds
belonging to the same complex and using different clump identification
algorithms.

\section{Conclusions}
The \ceiot\ map of the Chamaeleon~{\sc i} cloud with an angular 
resolution of
$\sim 1\arcmin$ shows a network of filaments with the likeness of some
recent simulations of molecular cloud fragmentation driven by large
scale turbulence (e.g.  \cite{klessen2001}). These filaments seem to
have, however, preferential directions, which change from N-S in the
north to NE-SW in the other parts. 'Cores' identified in the previous
NANTEN map (\cite{mizuno1999}) break up into a number of
substructures, which we call 'clumps'.  We have analyzed the small
scale structure using the 'Clumpfind' routine (\cite{williams1994})
and identified about 70 separate clumps in the $(x,y,v)$ data cube.
The RMS noise level of these data limits the detectable clump spectrum
to masses above $0.5 \Msun$.
The inspection of the clump masses and kinematics shows that none of
these clumps is a gravitationally bound entity, unless a large
fraction of the mass is hidden from \ceio\ observations because of
depletion.  However, a low external pressure of about $10^{-12}$ Pa
($P/k \sim 7\,10^4$ cm$^{-3}$K) would be sufficient to bring some 40
clumps to a balance.  This kind of pressure could be caused by
turbulent interclump medium ($n \sim 10^3$ \percc) with a velocity
dispersion of $\sigma_v = 0.5$ \kmps \ or a magnetic field with a
strength of 16 $\mu$G.

Using the analytical model for turbulent fragmentation presented in
\cite{padoan1997} and \cite{padoan2002} we derived a theoretical mass
spectrum for clumps in virial equilibrium by help of external
pressure. It turned out that large turbulent Mach numbers are required
to produce a mass spectrum peaking at low masses as observed, and that
corresponding turbulent velocity dispersion clearly exceeds the
interclump velocity dispersion quoted above. A possible reason for the
discrepancy is that the turbulence in the cloud has decayed since the
hypothetical equilibrium clumps formed. Alternatively all
condensations seen in the \ceio\ map are just transient structures. In
any case, gravitationally bound protostellar cores, such as Cha-MMS1 and
Cha-MMS2 probably are, represent subcondensations of \ceio\ 'clumps'.

Although the clump mass distribution may be understood in terms of
turbulent fragmentation models, one observational result is not easily
explainable by them, namely, the well-ordered magnetic field structure
in the cloud.  The magnetic field direction as determined from the
polarization position angles of highly reddened background stars is
perpendicular to the general orientation of dense filaments,
suggesting a magnetically controlled collapse (\cite{mcgregor1994}).

We consider it possible that different processes are taking effect on
different scales. Although large scale magnetic fields are likely to
have influenced the formation of the filaments, their fragmentation
into smaller clumps may have been dominated by chaotic motions.
In the sequel, we intend to perform direct comparisons between the
observational data (spectra and column density distributions) and
simulation results in order to determine the characteristics of
the turbulence and thereby gain better understanding of its r{\^o}le in
the cloud fragmentation.

\begin{acknowledgements} We thank Jonathan Williams for making 
the IDL version of the Clumpfind program publicly available. 
This project was supported by the Academy
 of Finland, grant Nos. 73727, 74854.
\end{acknowledgements}


\begin{thebibliography}{}

\bibitem[Ballesteros-Paredes \& Mac Low 2002]{ballesteros2002}
         Ballesteros-Paredes, J. \& Mac Low, M.-M.\ 2002, \apj, 570, 734


\bibitem[Boulanger et al. 1990]{boulanger1990}
 Boulanger, F., Falgarone, E., Puget, J.~L.\&  Helou, G. \ 1990, \apj, 364, 136

\bibitem[Boulanger et al. 1998]{boulanger1998} 
Boulanger, F., Bronfman, L., Dame, T.~M. \&  Thaddeus, P.\ 1998, \aap, 332, 
273 

\bibitem[Cambr\'esy et al. 1997]{cambresy1997} Cambr\'esy, L., Epchtein, N., 
Copet, E., de  Batz, B., Kimeswenger, S., Le Bertre, T., Rouan, D. \&
Tiph\'ene, D., 1997, \aap 324, L5 

\bibitem[Cambr\'esy et al. 1998]{cambresy1998} Cambr\'esy, L., Copet, E., 
Epchtein, N., de  Batz, B., Borsenberger, J., Fouque, P.,  Kimeswenger, S., 
\& Tiph\'ene, D., 1997, \aap 338, 997 

\bibitem[Carpenter et al. 2002]{carpenter2002} Carpenter, J.~M.,
Hillenbrand, L.~A., Skrutskie, M.~F., \& Meyer, M.~R.\ 2002, \aj, 124,
1001

\bibitem[Caselli et al. 1999]{caselli1999} Caselli, P., Walmsley, 
C.~M., Tafalla, M., Dore, L., \& Myers, P.~C.\ 1999, \apjl, 523, L165 

\bibitem[Cederblad 1946]{cederblad1946} Cederblad, S.\ 1946, 
Meddelanden fr{\aa}n Lunds Astronomiska Observatorium, Serie II, 119, 1 

\bibitem[Frerking et al. 1982]{frerking1982} Frerking, 
M.~A., Langer, W.~D., \& Wilson, R.~W.\ 1982, \apj, 262, 590 

\bibitem[Gahm et al. 2002]{gahm2002} Gahm, G.~F., Lehtinen, K., 
Carlqvist, P., Harju, J., Juvela, M. \& Mattila, K.\ 2002, 
\aap, 389, 577 

\bibitem[G\'{o}mez \& Kenyon 2001]{gomez2001}G\'{o}mez, M. \& Kenyon, S.\ J.\  
2001,  \aj 121,   974

\bibitem[Goodman et al. 1992]{goodman1992} 
Goodman, A.~A., Jones, T.~J., Lada, E.~A., \& Myers, P.~C.\ 1992, \apj, 
399, 108 

\bibitem[Haikala et al. 1998]{haikala1998} 
Haikala, L.~K., Mattila, K., Lehtinen, K. \& Lemke, D.\ 1998, 
ASP Conf.~Ser.~132: Star Formation with the Infrared Space Observatory, 
147 

\bibitem[Harjunp\"a \"a \& Mattila 1996]{harjunpaa1996} Harjunp\"a\"a, 
P.~\& Mattila, K.\ 1996, \aap, 305, 920 


\bibitem[Hayakawa et al. 1999]{hayakawa1999} Hayakawa, T., Mizuno, 
A., Onishi, T., Hara, A., Yamaguchi, R. \& Fukui, Y.\ 1999, \pasj, 51, 919 

\bibitem[Jones et al. 1985]{jones1985} Jones, T.~J., Hyland, 
A.~R., Harvey, P.~M., Wilking, B.~A. \& Joy, M.\ 1985, \aj, 90, 1191 

\bibitem[Kenyon \& G{\' o}mez 2001]{kenyon2001} Kenyon, S.~J.~\& 
G{\' o}mez, M.\ 2001, \aj, 121, 2673 

\bibitem[Klessen 2001]{klessen2001} Klessen R.S. \ 2001, \apj, 556,
837

\bibitem[Klessen 2003]{klessen2003} Klessen R.S. \ 2003, in Reviews in 
Modern Astronomy, Vol. 16, astro-ph/0301381

\bibitem[Knude \& H{\o}g 1998]{knude1998} Knude, J. \& H{\o}g, E.\ 
1998, \aap, 338, 897 


\bibitem[Larson 1979]{larson1979} Larson, R.~B.\ 1979, \mnras, 
186, 479 

\bibitem[Larson 1981]{larson1981} Larson, R.~B.\ 1981, \mnras, 
194, 809 

\bibitem[Lehtinen et al. 2001]{lehtinen2001} 
Lehtinen, K., Haikala, L.~K., Mattila, K. \& Lemke, D.\ 2001, \aap, 367, 
311 

\bibitem[Lehtinen et al. 2003]{lehtinen2003} 
Lehtinen, K., Harju, J., Kontinen, S. \& Higdon, J.L.\ 2003, \aap, 401, 1017 

\bibitem[Mattila et al. 1989]{mattila1989}
Mattila, K., Liljestr{\"o}m, T. \& Toriseva, M.\ 1989, Proc. ESO Workshop on 
Low Mass Star Formation and Pre-main Sequence Objects, 153 

\bibitem[McGregor et al. 1994]{mcgregor1994} 
McGregor, P.~J., Harrison, T.~E., Hough, J.~H., \& Bailey, J.~A.\ 1994, 
\mnras, 267, 755 

\bibitem[Mizuno et al. 1998]{mizuno1998} Mizuno, A.,  Hayakawa, T,. 
Yamaguchi, N., et al. \ 1998, \apj, 507, 83

\bibitem[Mizuno et al. 1999]{mizuno1999} Mizuno, A. Hayakawa, T.,
Tachihara, K., et al.\ 1999, \pasj, 51, 859

\bibitem[Mizuno et al. 2001]{mizuno2001}
Mizuno, A., Yamaguchi, R., Tachihara, K., et al. \ 2001, \pasj, 53, 1071

\bibitem[Myers et al. 1991]{myers1991} Myers, P.~C., 
Ladd, E.~F., \& Fuller, G.~A.\ 1991, \apjl, 372, L95 

\bibitem[Padoan \& Nordlund 2002]{padoan2002} Padoan, P. \& Nordlund,
         {\AA}. \ 2002, \apj, 576, 870

\bibitem[Padoan et al. 1997]{padoan1997} Padoan, P., Nordlund, {\AA}. 
         \& Jones, B.J.T.  \ 1997, \mnras, 288, 145

\bibitem[Persi et al. 1999]{persi1999} Persi, P., Marenzi, 
A.~R., Kaas, A.~A., Olofsson, G., Nordh, L., \& Roth, M.\ 1999, \aj, 117, 
439 

\bibitem[Persi et al. 2000]{persi2000} Persi, P., 
Marenzi, A.R. \& Olosson, G. \ 2000, \aap, 357, 219 

\bibitem[Persi et al. 2001]{persi2001}
Persi P., Marenzi A.R., G\'omez M. \& Olofsson G.\ 2001, \aap, 376, 907

\bibitem[Prusti et al. 1991]{prusti1991}
Prusti, T., Clark, F.O., Whittet, D.C.B., Laureijs, R.J. \& 
Zhang, C.Y. \ 1991, \mnras  251, 303 

\bibitem[Schwartz 1991]{schwartz1991} Schwartz, R. D. \ 1991, 
Low Mass Star Formation in Southern Clouds, ESO Scientific Report
No.11, 1991, ed. B. Reipurth, p. 93

\bibitem[Reipurth et al. 1996]{reipurth1996} Reipurth, 
B., Nyman, L-{\AA}. \& Chini, R.\ 1996, \aap, 314, 258 

\bibitem[Tachihara et al. 2002]{tachihara2002} 
Tachihara, K., Onishi, T., Mizuno, A. \& Fukui, Y.\ 2002, \aap, 385, 909 

\bibitem[Toriseva \& Mattila 1985]{toriseva1985} Toriseva, M. \& Mattila,
K.\  1985, \aap, 153, 207 

\bibitem[Toriseva et al. 1985]{toriseva1985a} Toriseva, M, H{\" o}glund B. \& 
Mattila, K.\  1985, Rev. Mexicana Astron. Astrof., 10 135

\bibitem[Toriseva et al. 1990]{toriseva1990} 
Toriseva, M., Mattila, K., \& Bronfman, L.\ 1990, \apss, 171, 219 

\bibitem[T{\' o}th et al. 2000]{toth2000} T{\' o}th, L.~V., 
Hotzel, S., Krause, O., Lehtinen, K., Lemke, D., Mattila, K., Stickel, M.
\& Laureijs, R.~J.\ 2000, \aap, 364, 769 

\bibitem[Whittet et al. 1994]{whittet1994} Whittet, D.~C.~B., 
Gerakines, P.~A., Carkner, et al. \ 1994, \mnras, 268, 1 

\bibitem[Williams et al. 1994]{williams1994} Williams, 
J.~P., de Geus, E.~J. \& Blitz, L.\ 1994, \apj, 428, 693 

\bibitem[Wilson \& Rood 1994]{wilson1994} Wilson, T.~L. \& Rood, 
R.\ 1994, \araa, 32, 191 

\end{thebibliography}
\end{document}